\documentclass[fleqn,10pt]{wlscirep}
\usepackage{float}
\usepackage{comment}

\begin{document}

\title{Subgap dynamics of double quantum dot coupled between superconducting and normal leads}

\author[1,*]{B. Baran}
\author[1]{R. Taranko}
\author[1,**]{T. Doma\'{n}ski}
\affil[1]{Institute of Physics, M. Curie-Sk\l{}odowska University, 20-031 Lublin, Poland}
\affil[*]{bartlobaran@kft.umcs.lublin.pl}
\affil[**]{doman@kft.umcs.lublin.pl}

\date{\today}

\begin{abstract}
Dynamical processes induced by the external time-dependent fields can provide valuable insight into the characteristic energy scales of a given physical system. We investigate them here in a nanoscopic heterostructure, consisting of the double quantum dot coupled in series to the superconducting and the metallic reservoirs, analyzing its response to (i)~abrupt bias voltage applied across the junction, (ii)~sudden change of the energy levels, and imposed by (iii)~their periodic driving. We explore subgap properties of this setup which are strictly related to the in-gap quasiparticles and discuss their signatures manifested in the time-dependent charge currents. The characteristic multi-mode oscillations, their beating patters and photon-assisted harmonics reveal a rich spectrum of dynamical features that might be important for designing the superconducting qubits.
\end{abstract}

\maketitle

The double quantum dots embedded on interfaces between various external leads have been proposed 
for possible spin \cite{Nowack-2007} and spin-orbit quantum bits \cite{Nadj-Perge-2010}. Specifically, 
the superconducting qubits \cite{Antonov-2020} have been considered as promising candidates,  
making use of the bound states formed inside the pairing gap \cite{Nazarov-2010}. Their implementations could protect 
the parity of Cooper pairs on proximitized superconducting nonoscopic islands \cite{Marcus-2020}. 
Further perspectives for the proximitized double quantum dots appeared with 
the topological superconductors \cite{Flensberg-2012}, where the zero energy 
in-gap modes are protected by symmetry reasons. These Majorana-type quasiparticles could be 
used for constructing the charge qubit in a transmission line resonator (transmon) 
\cite{Hyart-2013} and may be incorporated in the gate tunable superconducting qubits (gatemons) 
\cite{Aguado-2020}. Readout by means of a switching-event measurement using the attached 
superconducting quantum interference devices has revealed quantum-state oscillations with 
sufficiently high fidelity \cite{Chiorescu-2003}, that seems appealing  for realization of quantum computing.

So far the static properties of in-gap bound states have been throughly investigated for 
the single and multiple quantum dots \cite{balatsky.vekhter.06,Rodero-11} and recently also 
for nanoscopic length atomic chains, semiconducting nanowires, and magnetic islands proximitized 
to bulk superconductors  \cite{Prada-2020}. Their particular realizations in the double quantum 
dots (DQDs) have been experimentally probed by the tunneling spectroscopy, using InAs 
\cite{Sherman.2017,Grove_Rasmussen.2018,Estrada_Saldana.2018,Estrada_Saldana.arxiv2018,Paaske-2020,Frolov-2021}, 
InSb \cite{Su.2017}, Ge/Si \cite{Zarassi.2017} and carbon nanotubes \cite{Cleuziou.2006,Pillet.2013} 
and by the scanning tunneling microscopy applied to various di-molecules deposited on 
superconducting substrates \cite{Bauer-2013,Ruby.2018,Franke-2018,Choi.2018,Kezilebieke.2019}. 
Rich properties of such in-gap bound states of the DQDs have been analyzed theoretically  
by a number groups \cite{Choi-2000,Zhu-2002,Tanaka.2010,Zitko.2010,Konig.2010,Rodero-11,
Droste.2012,Grifoni.2013,Brunetti-2013,Sothmann-2014,Meng.2015,Zitko-2015,Su.2017,Glodzik.2017,
Frolov-2018,Scherubl_2019,Wang-2019,Leijnse.2019}. 
Major features of two quantum dots coupled in series to the superconducting lead(s) originate
from the ground state configuration which can vary its even-odd parity, depending on: the energy 
levels, hybridization with the external reservoirs, the inter-dot coupling, and the Coulomb 
potential \cite{Tanaka.2010,Zitko-2015}. Such parity changes are corroborated by crossings of 
the in-gap bound states and can be empirically detected by discontinuities of the Josephson 
current in S-DQD-S junctions \cite{Grove_Rasmussen.2018,Estrada_Saldana.2018,Estrada_Saldana.arxiv2018} 
or the subgap Andreev current in N-DQD-S junctions \cite{Grove_Rasmussen.2018,Su.2017,Frolov-2021}. 
The resulting zero-bias conductance as a function the quantum dot levels (tunable by the 
plunger gates) resembles a honeycomb structure \cite{Grove_Rasmussen.2018,Estrada_Saldana.2018,
Estrada_Saldana.arxiv2018} instead of a diamond shape, typical for the single quantum dot junctions.
Influence of the coupling to external reservoirs is also meaningful. For instance in a regime of the strong 
coupling to superconducting lead(s) the spin of quantum dots would be screened \cite{Grove_Rasmussen.2018}.
In general, various arrangements of two quantum dots enable realization of the on-dot and inter-dot 
electron pairing, affecting the measurable charge transport properties \cite{Sothmann-2014}.  
In particular, for the singly occupied quantum dots (what can be assured by appropriate 
gating) the superconducting proximity effect could be blocked. Such triplet blockade effect has 
been recently reported in S-DQD-S \cite{Paaske-2020} and N-DQD-S \cite{Frolov-2021} nanostructures. 
As regards the Coulomb potential, its influence is indirectly manifested through the singlet-doublet 
transitions (related to variations between the even-odd occupancies of the quantum dots 
\cite{Paaske-2020,Frolov-2021}) and, under specific conditions,  can lead to the subgap 
Kondo effect \cite{Tanaka.2010,Pillet.2013,Bauer-2013,Zitko-2015,Novotny_etal-2021}. 

\begin{figure}
\centerline{\includegraphics[width=0.6\linewidth]{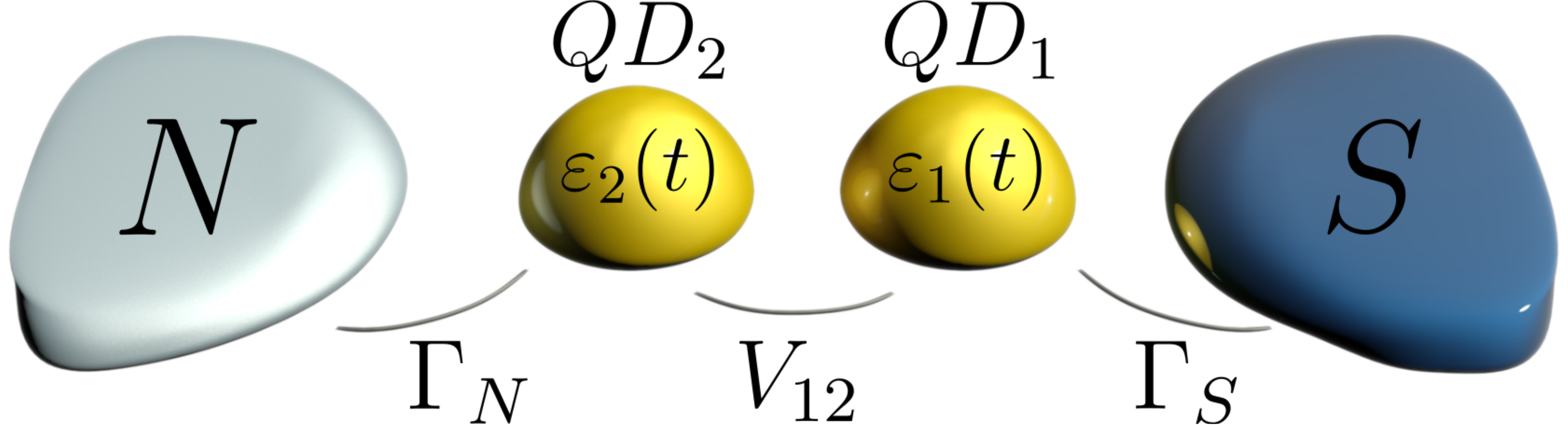}}
\caption{{\bf Schematics.} Two quantum dots (QD$_{1}$ and QD$_{2}$) coupled in series between  
the superconducting (S) and normal (N) metallic reservoirs whose energy levels $\varepsilon_{i\sigma}(t)$
could be varied by the external gate potential. We also consider dynamical phenomena driven by the 
time-dependent bias voltage imposed between the external leads.}
\label{fig:model}
\end{figure}

To our knowledge, however, the dynamical signatures of proximitized DQDs have not been investigated yet. 
Such dynamics could be important for designing future operations on the superconducting qubits, thefore
we analyze here various time-dependent observables of the setup, comprising two quantum dots arranged in series between the superconducting and normal metallic electrodes (Fig.~\ref{fig:model}). We inspect response of this heterostructure to several  types of external perturbations, leading either to a melting \cite{Essler-2020} or buildup \cite{Taranko-2019} of the electron pairing. For specific discussion we consider (i) abrupt detuning of the chemical potentials by the source-drain voltage, (ii) quench of the quantum dot energy levels, and (iii) their periodic driving. The latter effect has been recently achieved experimentally in the microwave-assisted tunneling  via the single quantum dot in the Josephson-type junctions \cite{Kot-2020,Franke-2020,vonOppen-2020}, but similar measurements should be  feasible using  N-DQD-S heterostructures as well.
Our calculations of the time-dependent electron occupancy and charge currents reveal the damped quantum oscillations whose frequencies coincide with the energies of in-gap bound sates. We inspect their nature and determine the characteristic time/energy scales, focusing on the limit of large superconductor gap, $\Delta=\infty$, and assuming the strongly asymmetric couplings, $\Gamma_{S}\gg\Gamma_{N}$. Under stationary condictions it has been shown for the single \cite{Tanaka_2007,Oguri_2012} and for the double quantum dot heterostructures \cite{Pokorny_2020} that
$\Delta \rightarrow \infty$ results do especially well and rather unexpectedly fit the results for systems with the finite pairing gap.
We show that periodic driving imposed on the quantum dot levels, $\varepsilon_{i\sigma}(t)$, induces the oscillating currents whose conductance (averaged over the period) has a structure  reminiscent of the Floquet systems. Dynamical properties studied in this work could be realized experimentally by applying either {\em dc} or {\em ac} external potentials.

\section*{Results}

We start by discussing the microscopic model of our setup (Fig.~\ref{fig:model}) and next present the numerical results obtained for three types of the quantum quench protocols. On this basis we infer the typical time-scales, characterizing in-gap bound states that would be useful for designing future operations on the Andreev qubits. In section {\em Methods} we present the eigenstates and eigenvalues for the case $\Gamma_{N}=0$ and provide some details about the computational techniques for N-DQD-S setup.

\subsection*{\label{sec:level2} Model and formalism}
%
Our heterostructure, consisting of the quantum dots  QD$_{i}$ ($i=1,2$) placed in linear  configuration between the normal (N) and superconducting (S) leads, can be described by the following Hamiltonian
\begin{equation}
\hat{H} = \hat{H}_{S}+\hat{H}_{S-QD_{1}}+\hat{H}_{DQD}+ \hat{H}_{N-QD_{2}}+\hat{H}_{N} .
\label{eq: 1}
\end{equation}
We treat the normal lead as free fermion gas $\hat{H}_{N}=\sum_{\textbf{k}\sigma}\xi_{N\textbf{k}\sigma}\hat{c}^{\dagger}_{N\textbf{k}\sigma}\hat{c}_{N\textbf{k}\sigma}$, where  $\hat{c}^{\dagger}_{N\textbf{k}\sigma}$ ($\hat{c}_{N\textbf{k}\sigma}$) is the creation (annihilation) operator of itinerant electron with the momentum $\textbf{k}$ and spin $\sigma$ whose energy $\xi_{N\textbf{k}\sigma}=\varepsilon_{N\textbf{k}\sigma}-\mu_{N}$ is measured with respect to the chemical potential $\mu_{N}$. The superconducting lead is assumed in the standard BCS form $\hat{H}_{S}=\sum_{\textbf{q}\sigma}\xi_{S\textbf{q}\sigma}\hat{c}^{\dagger}_{S\textbf{q}\sigma}\hat{c}_{S\textbf{q}\sigma}-\sum_{\textbf{q}}(\Delta_{SC}\hat{c}^{\dagger}_{S\textbf{q}\uparrow}\hat{c}^{\dagger}_{S\textbf{q}\downarrow}+h.c.)$, where $\Delta_{SC}$ stands for the isotropic pairing gap. 
The double quantum dot part is modeled by the single-level localized states
\begin{equation}
\hat{H}_{DQD}= \sum_{i\sigma}\varepsilon_{i\sigma}\hat{c}^{\dagger}_{i\sigma}\hat{c}_{i\sigma}+\sum_{\sigma} \left( V_{12} \hat{c}^{\dagger}_{1\sigma}\hat{c}_{2\sigma}+\mbox{\rm h.c.} \right),
\label{eq: 2}
\end{equation}
where $\hat{c}^{\dagger}_{i\sigma}$ ($\hat{c}_{i\sigma}$) is the creation (annihilation) operator of electron at $i$-th quantum dot, $\varepsilon_{i\sigma}$ denote for the energy levels, and $V_{12}$ is the interdot coupling. The quantum dots are hybridized with the external reservoirs via
$\hat{H}_{N-QD_{2}}= \sum_{\textbf{k}\sigma} \left(  V_{N\textbf{k}} \hat{c}^{\dagger}_{N\textbf{k}\sigma}\hat{c}_{2\sigma}+\mbox{\rm h.c.} \right)$ and
$\hat{H}_{S-QD_{1}} = \sum_{\textbf{q}\sigma} \left( V_{S\textbf{q}} \hat{c}^{\dagger}_{S\textbf{q}\sigma}\hat{c}_{1\sigma}+\mbox{\rm h.c.} \right)$,  
where $V_{N\textbf{k}}$ ($V_{S\textbf{q}}$) denotes the coupling to normal (superconducting) lead.  

We restrict our considerations to the wide-band limit, assuming the constant (energy-independent) auxiliary couplings $\Gamma_{N/S}=2\pi\sum_{\textbf{k}/\textbf{q}}|V_{N\textbf{k}/S\textbf{q}}|^{2}\delta(\varepsilon-\epsilon_{N\textbf{k}/S\textbf{q} \sigma})$. We also treat the pairing gap $\Delta_{SC}$ as the largest energy scale, focusing on dynamical processes solely inside in the subgap regime. 
In the limit of infinite $|\Delta|$ the selfenergy of the Nambu-matrix Green's function becomes static and the value $\Gamma_{S}/2$ appearing in the off-diagonal terms can be interpreted as the proximity induced pairing potential.
The resulting low-energy physics can be described by \cite{Oguri-2004}
\begin{equation}
\hat{H}_{S}+\hat{H}_{S-QD_{1}} \approx  \frac{\Gamma_{S}}{2} \left( \hat{c}_{1\downarrow}^{\dagger}\hat{c}_{1\uparrow}^{\dagger}
+ \hat{c}_{1\uparrow}\hat{c}_{1\downarrow} \right) . 
\label{eq: 4}
\end{equation}

In what follows we discuss the time-dependent charge currents $j_{N\sigma}(t)$, $j_{S\sigma}(t)$ and occupancies of the quantum dots imposed by the following types of quantum quenches: 
(i)~abrupt bias potential $V_{sd}=\mu_{N}-\mu_{S}$ applied between $N$ and $S$ electrodes,
(ii)~sudden change of the energy levels $\varepsilon_{i\sigma}$ due to the gate potential $V_{g}$, and
(iii)~periodic driving of the quantum dot levels with a given amplitude and frequency.
Expectation values of the physical observables are computed numerically, solving a closed set of the differential equations for appropriate correlation functions (see {\em Methods}).
The charge current $j_{N\sigma}(t)$ flowing between the normal lead and QD$_{2}$ can be derived from the time-dependent number of electrons in the normal lead. For $\varepsilon_{N\textbf{k}\sigma}(t)=\varepsilon_{N\textbf{k}\sigma}$ this current is formally given by \cite{Taranko-2018}  
\begin{eqnarray}
j_{N\sigma}(t)&=&2\textrm{Im}\left( \sum_{\textbf{k}}V_{N\textbf{k}}\exp(-i\varepsilon_{N\textbf{k}\sigma}t)\langle \hat{c}^{\dagger}_{2\sigma}(t)\hat{c}_{N\textbf{k}\sigma}(0)\rangle \right) 
-\Gamma_{N} \; n_{2\sigma}(t),
\label{j_N}
\end{eqnarray}
were $\langle \dots \rangle$ denotes the quantum statistical averaging and  $\langle n_{i\sigma}(t) \rangle\equiv \hat{n}_{i\sigma}(t)$. The interdot charge flow $j_{12\sigma}(t)$ is expressed as
\begin{equation}
j_{12\sigma}(t)=-\textrm{Im}\left( V_{12}\langle \hat{c}^{\dagger}_{1\sigma}(t)\hat{c}_{2\sigma}(t)\rangle\right)
\label{eq: 6}
\end{equation}
whereas the current $j_{S\sigma}(t)$ flowing from the superconducting lead to QD$_{1}$ can be obtained from the charge conservation law $\frac{dn_{1\sigma}(t)}{dt}=j_{12\sigma}(t)+j_{S\sigma}(t)$.
Using equation (\ref{j_N}) for the current $j_{N\sigma}$ we can define its time-dependent differential conductance $G_{N\sigma}(V_{sd},t)=\frac{d}{dV_{sd}}j_{N\sigma}(t)$ as a function of the source-drain voltage $V_{sd}$. Peaks appearing in the dependence of $G_{N\sigma}(V_{sd},t)$ against $V_{sd}$ can be interpreted as the excitation energies between eigenstates, comprising even and odd number of electrons (dubbed the Andreev bound states). Upon approaching the steady limit, $t\rightarrow\infty$, they emerge in the uncorrelated system at energies $E=\pm\frac{1}{2}\left(\sqrt{4V^{2}_{12}+\Gamma^{2}_{S}/2}\pm\frac{\Gamma_{S}}{2}\right)$  (see {\em Methods}) and acquire a finite broadening caused by the relaxation processes on continuous spectrum of the normal lead.

In practical realizations of such N-DQD-S heterostructure (Fig.~\ref{fig:model}) one should also take into account the Coulomb repulsion between electrons,  $\sum_{i=1,2}U_{i}n_{i\uparrow}n_{i\downarrow}$, competing with the proximity-induced electron pairing and thereby affecting the bound states. Some aspects of the correlations effects have been previously studied under the stationary conditions for this heterostructure by the numerical renormalization group method \cite{Tanaka.2010}.
Here we shall address the post-quench dynamics, treating the electron-electron interactions within the Hartree-Fock-Bogoliubov decoupling scheme 
\begin{equation}
\hat{n}_{i\uparrow}\hat{n}_{i\downarrow}\simeq   \hat{n}_{i\uparrow}\langle \hat{n}_{i\downarrow} \rangle +\hat{n}_{i\downarrow}\langle \hat{n}_{i\uparrow} \rangle +\hat{c}^{\dagger}_{i\uparrow}\hat{c}^{\dagger}_{i\downarrow}\langle \hat{c}_{i\downarrow}\hat{c}_{i\uparrow} \rangle+\hat{c}_{i\downarrow}\hat{c}_{i\uparrow}\langle \hat{c}^{\dagger}_{i\uparrow}\hat{c}^{\dagger}_{i\downarrow} \rangle .
\label{HFB-approx}
\end{equation}
This approximation applied to the static case of the correlated quantum dot hybridized with superconducting lead(s)  can qualitatively describe the parity crossings and the energies of in-gap bound states \cite{Martin_Rodero_2012}. We use of this decoupling (\ref{HFB-approx}) to provide a preliminary insight into the complicated quench-driven dynamics of the interacting setup, which is effectively described by 
\begin{equation}
\hat{H}_{eff} \approx  \sum_{i,\sigma} \tilde{\varepsilon}_{i\sigma}(t)  \hat{c}_{i\sigma}^{\dagger} \hat{c}_{i\sigma} - \sum_{i} \left( \Delta_{i}(t)  \hat{c}_{i\uparrow}^{\dagger}  \hat{c}_{i\downarrow}^{\dagger} + \mbox{\rm h.c.} \right) +  \sum_{\sigma} \left( V_{12} \hat{c}_{1\sigma}^{\dagger} \hat{c}_{2\sigma} + \mbox{\rm h.c.}\right) + \sum_{\textbf{k},\sigma}\left(V_{N\textbf{k}}  \hat{c}_{N\textbf{k}\sigma}^{\dagger} \hat{c}_{2\sigma}  + \mbox{\rm h.c.}\right)  +\sum_{\textbf{k}\sigma}\xi_{N\textbf{k}\sigma}\hat{c}^{\dagger}_{N\textbf{k}\sigma}\hat{c}_{N\textbf{k}\sigma} 
\label{effective_model} 
\end{equation}
with the renormalized energy levels $\tilde{\varepsilon}_{i\sigma}(t)=\varepsilon_{i\sigma}(t)+U_{i} n_{i\sigma}(t)$ and the effective on-dot pairings $\Delta_{1}(t)=\frac{\Gamma_{S}}{2}-U_{1}\langle \hat{c}_{1\downarrow}(t)\hat{c}_{1\uparrow}(t)\rangle$,
$\Delta_{2}(t)=-U_{2}\langle \hat{c}_{2\downarrow}(t)\hat{c}_{2\uparrow}(t)\rangle$.
Such mean-field approximation might be reliable at least for the weak interaction case. More subtle analysis, including the Kondo effect of the strongly correlated system ($U_{i} \gg \Gamma_{S}$), is beyond a scope of this paper. We have done numerical calculations for  $U_{1}=U_{2} \equiv U$, considering $U/\Gamma_{S}=0.5$, $1$ and $1.5$, respectively. Technically we have adapted for this purpose the algorithm outlined in {\it Methods}, extending the previous study of the single dot superconducting junctions \cite{Taranko-2018,Taranko-2019}.

We use the convention $e=\hbar=1$, expressing the charge currents, time and frequency $\omega$ in units of $e\Gamma_{S}/\hbar$, $\hbar/\Gamma_{S}$ and  $\Gamma_{S}/\hbar$, respectively. In realistic experimental situations the value of $\Gamma_{S} \sim 200$ $\mu$eV would imply the following typical units of time $\sim 3.3$ psec, current $\sim 48$ nA and frequency $\sim 0.3$ THz. We assume the superconducting lead to be grounded, treating its chemical potential as the convenient reference level ($\mu_{S}=0$). Our calculations are performed for zero temperature.

\subsection*{Response to a bias voltage}

\begin{figure}[t!]
\centerline{\includegraphics[width=1\linewidth]{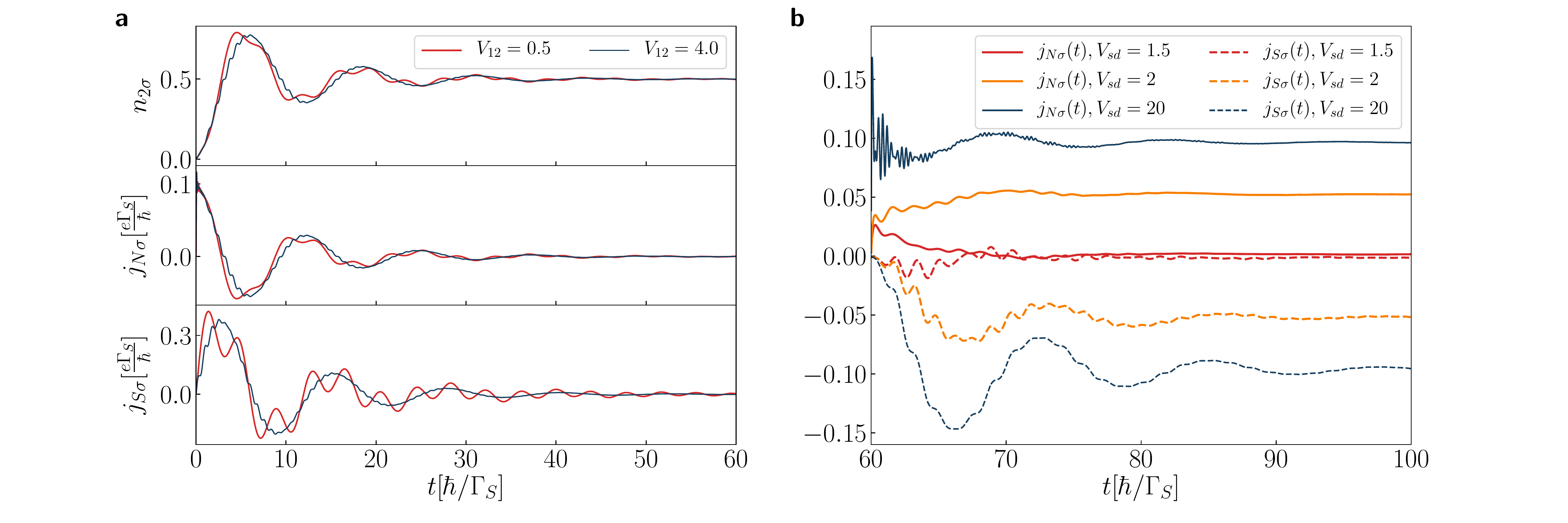}}
\caption{{\bf Transient and post-quench dynamics.} a) The time-dependent charge $n_{2\sigma}$ and transient currents $j_{S\sigma}$, $j_{N\sigma}$ obtained for  $V_{12}/\Gamma_{S}=0.5$, $4$, assuming the initially empty quantum dots. b) The post-quench currents $j_{S\sigma}$ and $j_{N\sigma}$ for $V_{12}/\Gamma_{S}=2$ after an abrupt biasing by the source-drain voltage $V_{sd}$ at $t=60$. Calculations have been done for $U=0$, $\varepsilon_{i\sigma}=0$, $\Gamma_{N}/\Gamma_{S}=0.2$.}
\label{fig: 2}
\end{figure}

For computational reasons it is convenient to assume that initially, at $t = 0$, the quantum dots are 
disconnected from both external reservoirs (see {\em Methods}). Figure \ref{fig: 2}a presents the transient currents $j_{N\sigma}(t)$ and $j_{S\sigma}(t)$ right after forming the N-DQD-S heterostructure.  In analogy to the previously discussed N-QD-S case \cite{Taranko-2018} 
such evolution to the stationary limit is achieved through a sequence of the damped quantum oscillations, whose 
frequencies coincide  with the energies of in-gap bound states. In particular, for $\varepsilon_{i\sigma}=0$ 
the period of such oscillations is equal to $T=4\pi/\Gamma_{S}$ and the relaxation processes (originating from the coupling 
$\Gamma_{N}$ of QD$_{2}$ to the metallic lead) impose the damping via exponential envelope function 
$e^{-t\Gamma_{N}/2}$. In practice, at times $t \geq 50$, the stationary state seems to be fairly well approached.

Let us turn to the dynamical response of N-DQD-S setup induced by its biasing, at $t=60$, when the chemical 
potentials are detuned by by source-drain voltage $\mu_{N}-\mu_{S}=V_{sd}$. Figure \ref{fig: 2}b presents the charge 
currents $j_{N\sigma}(t)$ and $j_{S\sigma}(t)$ obtained for $V_{12}/\Gamma_{S}=2$, assuming $V_{sd}/\Gamma_{S}=1.5$, $2$ and $20$, respectively. 
For the large bias voltage, $|V_{sd}| \gg V_{12}$, we observe emergence of the quantum beats  with the period $T_{B}=\pi/V_{12}$ superimposed with the higher frequency oscillations. Let us recall that charge transport is provided here solely by the anomalous particle-to-hole (Andreev) scattering, which is sensitive to the in-gap bound states. For the particular set of model parameters such in-gap bound states appear at energies $\pm \frac{1}{2}\sqrt{4V_{12}^{2}+\Gamma_{S}^{2}/4}\pm \Gamma_{S}/4$. 
It has been previously shown \cite{Taranko-2012} that the single quantum dot placed between both normal electrodes responds to a sudden external voltage by the coherent oscillations of the charge current with frequency  $\omega=|V_{sd}-\varepsilon_{dot}|$. In the present situation we should replace $\varepsilon_{dot}$ by the effective in-gap quasiparticle energies, at which the Andreev scattering is amplified.
We have four such in-gap bound states, therefore total current can be viewed as a superposition of sinusoidal waves, oscillating with the frequencies $\Omega_{1/2}=V_{sd}\pm\omega_{1}$ and $\Omega_{3/4}=V_{sd}\pm\omega_{2}$, where $\omega_{1/2}=V_{12}\pm\Gamma_{S}/4$. It can be effectively expressed as $\sum^{4}_{i=1}a_{i}e^{-\lambda_{i}t}\sin(\Omega_{i}t)$. 
Individual terms refer here to the damping processes with different parameters $\lambda_{i}$, whereas the coefficients $a_{i}$ control the contributions from these in-gap bound states. For the large bias $|V_{sd}| \gg V_{12}$ and $|V_{sd}| \gg \Gamma_{S}/4$ the quantum beats are superimposed with the faster oscillations. It can be shown  \cite{Taranko-2012} that such beating patterns depend on a ratio 
\begin{equation}
r=\frac{\omega_{1}+\omega_{2}}{|\omega_{1}-\omega_{2}|}=\frac{4V_{12}}{\Gamma_{S}}.
\end{equation}
For the case displayed in Fig.~\ref{fig: 2}b this ratio is $r=8$, therefore for $V_{sd}/\Gamma_{S}=20$ the repeated sequences of the beats with the periods $\frac{\pi}{4}, \frac{\pi}{2}, \frac{\pi}{2}, \frac{\pi}{2}, \frac{\pi}{2}, \frac{\pi}{2}, \frac{\pi}{2}, \frac{\pi}{2},  \frac{\pi}{4}$ appearing in the current $j_{N\sigma}(t)$  should be  observed.  For non-integer ratio $r$ the resulting beating pattern is more complicated with the different successive periods.
Figure \ref{fig: 2}b displays that for $V_{sd}/\Gamma_{S}=20$ the post-quench current $j_{N\sigma}(t)$ indeed exhibits the beats mainly with period $T_{B}=\pi/V_{12}$ superimposed with the faster oscillations, whose frequency is equal to $V_{sd}$. The steady limit current obtained for $V_{sd}/\Gamma_{S}=2$ is larger than for $V_{sd}/\Gamma_{S}=1.5$ because of the broader transport window involving all the in-gap bound states. 
We also notice that $j_{S\sigma}(t)$ substantially differs from $j_{N\sigma}(t)$, especially for the large bias $V_{sd}$. We assign this to the fact that DQD sandwiched between the external leads wash out small fluctuations of the current $j_{S\sigma}(t)$, enforcing the final damped oscillations with period $4\pi/\Gamma_{S}$.

\begin{figure}
\centering
\includegraphics[width=0.5\columnwidth]{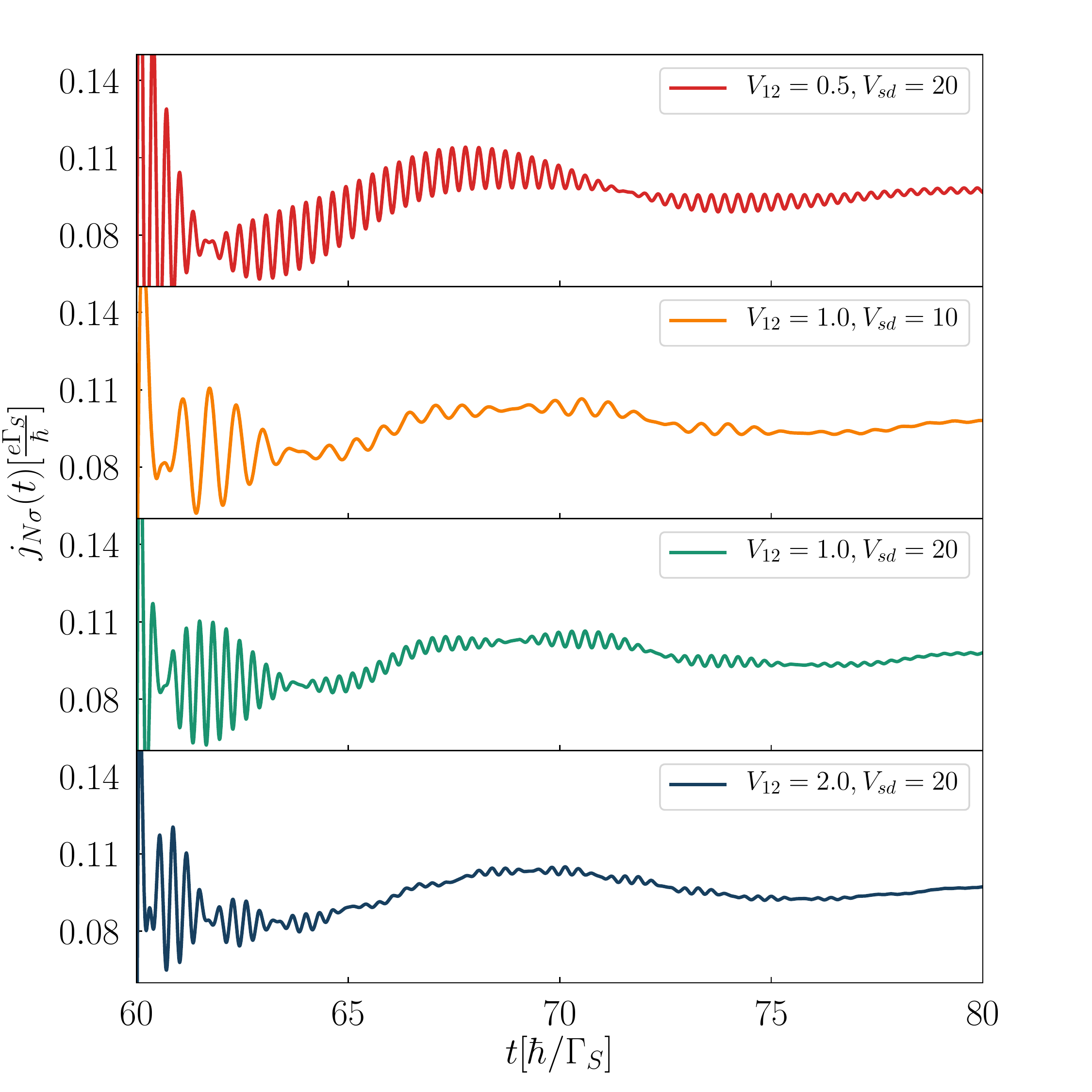}
\caption{{\bf Post-quench beating patterns.} The Andreev current $j_{N\sigma}(t)$ induced by abrupt biasing at $t=60$ for several values of the interdot coupling $V_{12}$ and $V_{sd}$ (in units of $\Gamma_{S}$), as indicated. We used the model parameters $U=0$, $\varepsilon_{i\sigma}=0$,  $\Gamma_{N}/\Gamma_{S}=0.2$.}
\label{fig: 3}
\end{figure}

Fig.\ref{fig: 3} shows the beating structure in the time-dependent current $j_{N\sigma}(t)$ after abrupt application of the bias voltage. These beats clearly depend on the interdot coupling $V_{12}$ via $T_{B}=\pi/V_{12}$. The beating structure is superimposed with oscillations whose frequency is also sensitive to the bias voltage. 
By measuring the period of such beating oscillations one could thus practically evaluate the inter-dot coupling $V_{12}=\pi/T_{B}$. For a realistic value $\Gamma_{S}\sim 200\mu eV$, and assuming $V_{12}/\Gamma_{S}=0.5$, $1$ and $2$ the beating period  would be $T_{B}\sim 21$, $10$ and $5$ picoseconds, respectively. This time-scale is currently attainable experimentally.
We have also performed similar calculations including the electron correlations (within the mean-field approximation assuming $\varepsilon_{i\sigma}=-U/2$) and found, to our surprise, that all conclusions concerning the frequencies and the beating patterns remain valid. 

\subsection*{Quench of energy levels}

Let us now consider the dynamics induced by a sequence of quantum quenches imposed on the energy levels $\varepsilon_{i\sigma}$. The first quench $\varepsilon_{i\sigma}\rightarrow \varepsilon_{i\sigma}+V_{g}$  is performed at $t_{1}=60$, safely after  N-DQD-S heterostructure achieves its stationary configuration. Later on, at time $t_{2}=120$, we rapidly change the energy levels back to their initial values $\varepsilon_{i\sigma}+V_{g} \rightarrow \varepsilon_{i\sigma}$. Such step-like change (reminiscent of the pump-and-probe techniques) could be practically driven by the external gate potential applied to DQDs.

\begin{figure}[h]
\centering
\includegraphics[width=0.5\columnwidth]{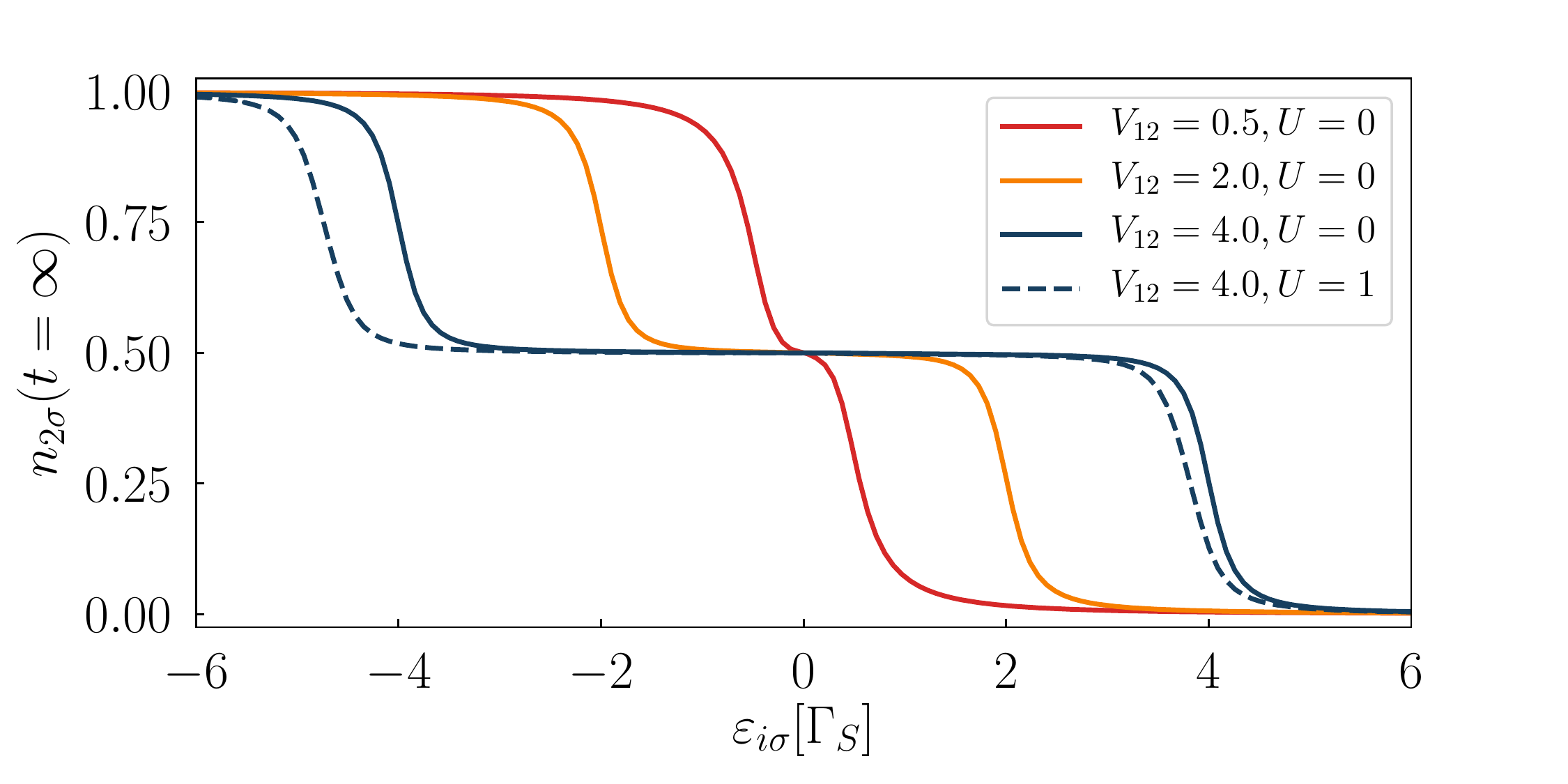}
\caption[width=1\textwidth]{{\bf Charge occupancy.} The stationary limit $(t=\infty)$  of the QD$_{2}$ occupancy as a function of the energy level $\varepsilon_{2\sigma}=\varepsilon_{1\sigma}$ determined for several interdot couplings $V_{12}$. The dashed line is calculated within the mean-field approximation for $U=1$. Other parameters: $V_{sd}=0$, $\Gamma_{N}=0.1$, $\Gamma_{S}=1$.}
\label{fig: 4}
\end{figure}

For understanding the dynamics of our setup it is helpful to inspect the stationary fillings of both quantum dots for various interdot couplings $V_{12}$.  Fig.~\ref{fig: 4} shows the occupancy of QD$_{2}$ (the neighbor of the normal lead) with respect to the energy level $\varepsilon_{2\sigma}$, assuming $\varepsilon_{1\sigma}=\varepsilon_{2\sigma}$ so that occupancies of both dots are nearly identical. We recognize three plateau regions, corresponding to $n_{i\sigma}\approx 1$, 0.5 and 0, respectively. 
We also notice, that a width of the half-filling region strongly depends on the inter-dot coupling $V_{12}$. The stationary occupancy $n_{2\sigma}$ changes from the nearly complete filling to half-filling or from the half-filled case to nearly empty QDs occur in a vicinity of $\varepsilon_{i\sigma}\approx\pm V_{12}$ where the in-gap bound states coincide with the chemical potential $\mu_{N}=\mu_{S}$ (here $V_{sd}=0$). 
Our numerical results obtained for various $V_{12}$ and $V_{g}$ indicate that the most prominent changes of the time-dependent observables occur for such quenches when the final value of the energy levels $\varepsilon_{i\sigma}$ coincide with the changeovers of $n_{i\sigma}(t=\infty)$ illustrated in Fig.\ref{fig: 4}. We have also checked that postquench evolution for different interdot couplings $V_{12}$ preserves the same universal properties, provided that the final value $\varepsilon_{i\sigma}$ corresponds to the tilted part of $n_{i\sigma}(t=\infty)$ curve.

\begin{figure*}[ht]
\centering
\includegraphics[width=1\textwidth]{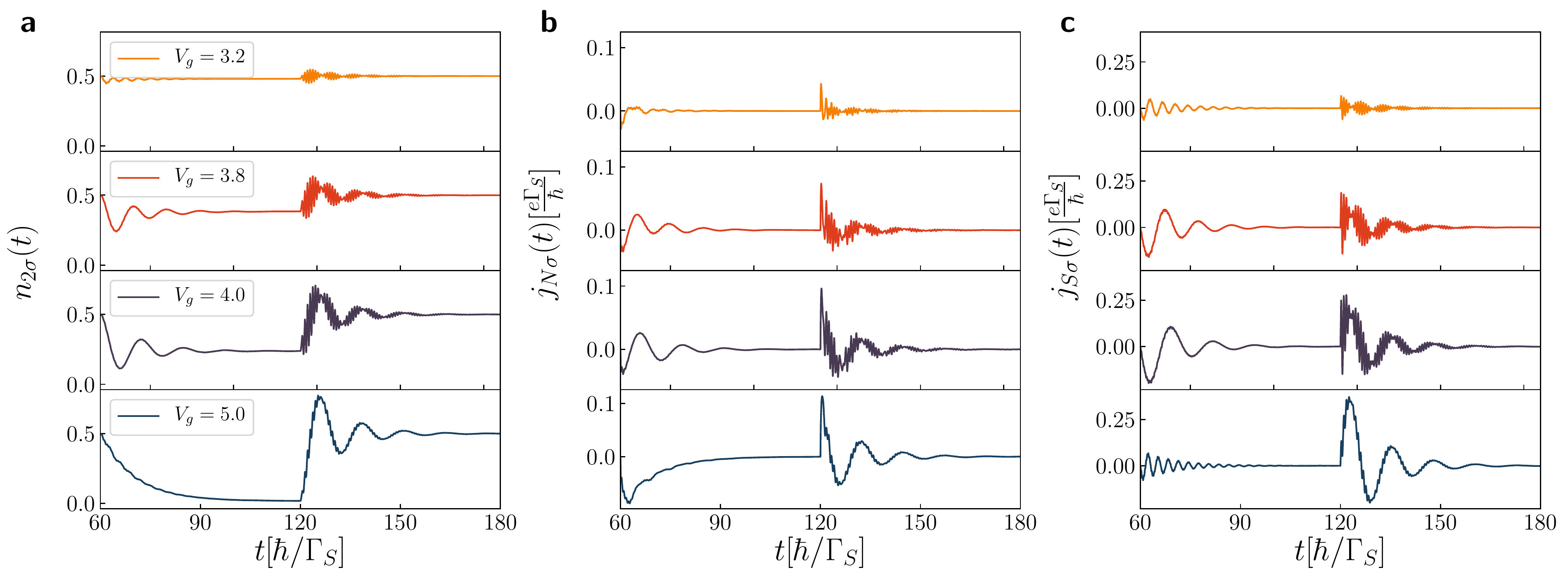}
\caption[width=1\textwidth]{{\bf Dynamics imposed by quench of energy levels.} The time-dependent occupancy $n_{2\sigma}(t)$ and the currents $j_{N\sigma}(t)$, $j_{S\sigma}(t)$ driven by the step-like variation of the energy levels $\varepsilon_{i\sigma}\rightarrow \varepsilon_{i\sigma}+V_{g}$, at $t=60$, and $\varepsilon_{i\sigma}+V_{g} \rightarrow \varepsilon_{i\sigma}$, at $t=120$. Results are obtained for $U=0$,  $V_{sd}=0$, $\varepsilon_{i\sigma}=0$, $\Gamma_{N}/\Gamma_{S}=0.2$, $V_{12}/\Gamma_{S}=4$, and several values of $V_{g}$ (in units of $\Gamma_{S}$) as indicated.}
\label{fig: 5}
\end{figure*}

Figure \ref{fig: 5} shows the time-dependent $n_{2\sigma}(t)$,  $j_{N\sigma}(t)$, and $j_{S\sigma}(t)$  after lifting the DQD energy levels, at $t=60$, and their return to initial values, at $t=120$, obtained for the strong interdot coupling, $V_{12}/\Gamma_{S}=4$. For $t\leq 60$ the considered N-DQD-S system is practically in its stationary state with the half-filled QDs and negligible currents $j_{N\sigma}(t)$, $j_{S\sigma}(t)$.
More specifically, we have chosen $V_{g}/\Gamma_{S}=3.2$, $3.8$, $4$, and $5$, respectively. Such values of $V_{g}$ correspond to the stationary occupancies equal to $\sim 0.48$, $\sim 0.4$, $\sim 0.25$ and $\sim 0.015$, respectively. 
Let us consider the postquench evolution corresponding to $V_{g}/\Gamma_{S}=3.2$, when the quantum dot level $\varepsilon_{i\sigma}$ coincides with the middle plateau (Fig.\ref{fig: 4}). The initial occupancy of QD$_{2}$ is $0.5$ and its stationary value after the first quench (at $t=60$) changes to $\sim 0.48$, therefore $n_{2\sigma}(t)$ exhibits only very small oscillations. Similarly, the charge currents $j_{N\sigma}$ and $j_{S\sigma}$ are negligible (see the upper curves in Fig.~\ref{fig: 5} for $t<120$). 
After the second quench (at $t=120$) the occupancy $n_{2\sigma} \sim 0.5$, albeit promptly after the quench we observe some transient phenomena with the beating structure (see the upper curve in Fig.~\ref{fig: 5} for $t>120$). This beating structure is more evident for the larger gate potentials $V_{g}/\Gamma_{S}=3.8$ and  $4$ (see Fig.~\ref{fig: 5}). We observe oscillations with the period $T=\pi/V_{12}$, giving rise to the beating structure with another period $2\pi/\Gamma_{S}$.
Upon increasing the gate potential to $V_{g}/\Gamma_{S}=5$ the time-depenence of $n_{2\sigma}$ after the first quench substantially changes in comparison with the previous cases. Instead of the damped oscillations we now observe an exponential decay, down to nearly zero. Evolution after the second quench is also different in comparison to the previous ones. We now observe the oscillations of $n_{2\sigma}$ and both currents with the period $T=2\pi/\Gamma_{S}$ without any beating structure. 
%
%
Concerning the time-dependent occupancies and currents calculated for $V_{12}/\Gamma_{S}\geq 1$, they preserve the qualitative properties discussed above. For the smaller interdot couplings $V_{12}$ (for instance $V_{12}/\Gamma_{S}=0.5$) the evolution after the first quench preserves all properties characterized for stronger $V_{12}$, but after the second quench we no longer observe the beating patterns, so that only oscillations with the period  $4\pi/\Gamma_{S}$ are present.

We have also performed calculations for the interacting system, assuming $U/\Gamma_{S}=1$. The stationary limit occupancy of QD$_{2}$ is shown by the dashed line in  Fig.\ref{fig: 4}. We can notice that the characteristic points, where the totally filled dot changes to the half-filling and another one where the half-filled dot changes to the empty configuration, are shifted in comparison to the noninteracting case. This effect is caused by rescaling of the in-gap states energies. In analogy to our considerations of uncorrelated system we have imposed such variations of the quantum dot levels by the gate potential $V_{g}$ which coincided with these characteristic points of $n_{i\sigma}(t=\infty)$. It turned out that postquench evolution revealed the same qualitative features in the time-dependent occupancy $n_{2\sigma}(t)$ and the charge currents as for $U=0$. For brevity, we hence skip such results.
%

\subsection*{Periodically driven energy levels}

We now discuss dynamical response of the N-DQD-S heterostructure driven by a periodic driving 
of the energy levels $\varepsilon_{i\sigma}(t)=A\sin(\omega t)$ which can be practically 
achieved by shining an infrared field on the quantum dots. We assume that amplitude $A$ and 
frequency $\omega$ of the oscillations are identical in both QDs. 
\begin{figure}
\centering
\includegraphics[width=1\columnwidth]{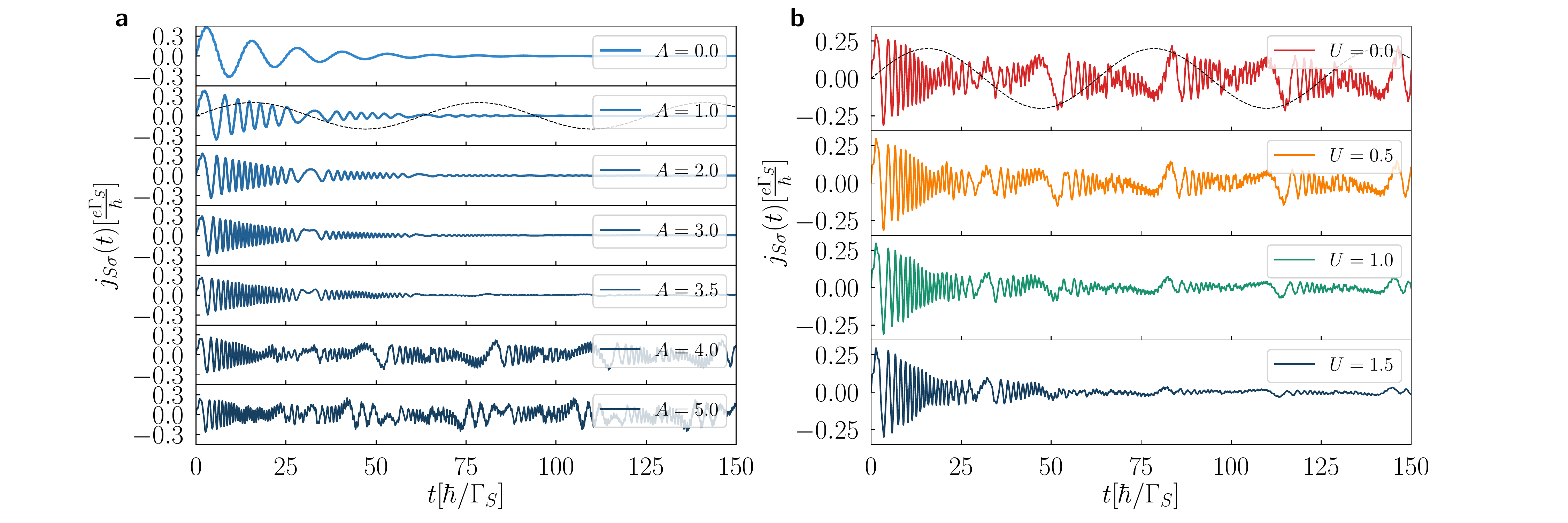}
\caption{{\bf Amplitude effect of periodic driving.} The current $j_{S\sigma}$ induced by the oscillating energy levels $\varepsilon_{i\sigma}(t)=A\sin(\omega t)$. Panel (a) presents the results obtained in uncorrelated system for $V_{12}/\Gamma_{S}=4$ and several amplitudes $A$. Panel (b) shows the mean-field results determined for $V_{12}/\Gamma_{S}=3$, $A/\Gamma_{S}=3$ and several values of the Coulomb potential $U$ (in units of $\Gamma_{S}$). We used the model parameters $V_{sd}=0$, $\omega=0.1/\Gamma_{S}$, $\Gamma_{N}/\Gamma_{S}=0.1$. The dashed lines illustrate profile of the oscillating energy levels (not in scale). } 
\label{fig: 6}
\end{figure}

Figure \ref{fig: 6} presents the time-dependent current $j_{S\sigma}(t)$ obtained  for $\omega/\Gamma_{S}=0.1$ and several  values of the amplitude $A$. The left (a) panel refers  to the uncorrelated case, $U=0$, and the right (b) panel to $U/\Gamma_{S}=1$, respectively. As a guide to eye we also display the transient current obtained for the static energy levels $\varepsilon_{i\sigma}=0$ (top panel in Fig.~\ref{eq: 6}a) with the characteristic damped oscillations whose period is equal to $4\pi/\Gamma_{S}$. Such current vanishes in the asymptotic limit $t\rightarrow\infty$ (here $V_{sd}=0$) and similar features, but with different profiles of the quantum oscillations, are observable for small amplitudes of the periodic driving as well. 
They are displayed for $V_{12}/\Gamma_{S}=4$ in Fig.~\ref{fig: 6}a. We notice that indeed the time-dependent currents asymptotically vanish for $A/\Gamma_{S}\leq 3.5$. This situation occurs whenever the amplitude $A$ does not exceed the energies of subgap quasiparticles. 
Such behavior can be contrasted with the larger amplitude driving (for instance $A/\Gamma_{S}=4$) when the current $j_{S\sigma}(t)$ is forced to flow back and forth all over the time. Periodicity is this behavior is a bit subtle and will be analyzed in more detail underneath. 

Figure \ref{fig: 6}b shows the current $j_{S\sigma}(t)$ of the correlated system (Coulomb potential $U_{1}=U_{2}=U$ is expressed in units of $\Gamma_{S}$) determined for $V_{12}/\Gamma_{S}=3$, $A/\Gamma_{S}=3$, and $V_{sd}=0$. We have chosen such  parameters to enforce the nonvanishing current, up to the asymptotic limit $t\rightarrow\infty$. 
The correlation effects are here quite evident. Upon increasing $U$ the magnitude of oscillating current $j_{S\sigma}(t)$ is gradually suppressed. Such effect can be partly assigned to shifting of the subgap quasiparticles to the higher energies
and partly to ongoing transfer of the spectral weights (this behavior is also discussed in next subsection).
In presence of the finite source-drain voltage $V_{sd}$ the time-dependent phenomena become even more complicated. Its seems, however, that under such highly non-equilibrium conditions the correlation effects become less important.

\begin{figure}[H]
\centering
\includegraphics[width=0.5\columnwidth]{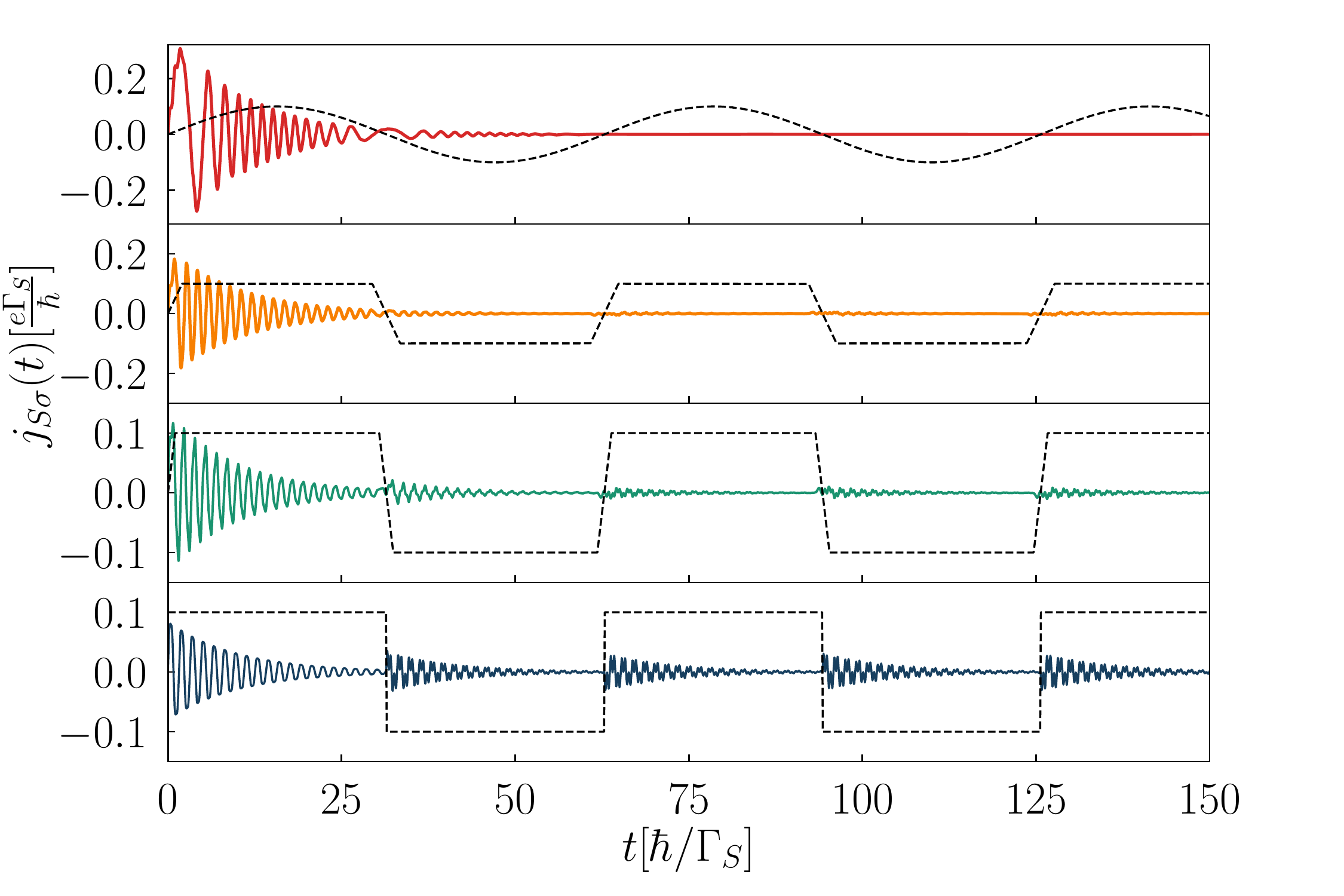}
\caption{{\bf Various profiles of periodic driving.} The time-dependent current $j_{S\sigma}$ (solid lines) obtained for several schemes of the periodic driving illustrated by dotted lines (not in scale). The results are obtained for $V_{12}/\Gamma_{S}=4$, $A/\Gamma_{S}=2$, $\omega/\Gamma_{S}=0.1$, using the model parameters  $U=0$, $\Gamma_{N}/\Gamma_{S}=0.2$, $V_{sd}=0$.}
\label{fig: 8}
\end{figure}

Finally we briefly investigate the transient currents imposed by different profiles of the periodically driven energy levels $\varepsilon_{i\sigma}(t)=\varepsilon_{i\sigma}(t+T)$ as depicted by the dashed lines in Fig.~\ref{fig: 8}. For all cases we have assumed the same amplitudes and frequencies. As the reference, the upper panel shows the case of the sinusoidally driven energy level.
It appears that abrupt (step-like) variations of QDs energy levels are followed by the damped oscillations of transient current $j_{S\sigma}(t)$ after each change of $\varepsilon_{i\sigma}$. Life-time of the resulting damped oscillations is shorter or comparable to the period of driving. For more smooth variation of $\varepsilon_{i\sigma}$ we can notice gradual suppression of the induced oscillations (see the second panel from the top of Fig.~\ref{fig: 8}).

\subsection*{Andreev conductance averaged over driving period}
\label{section: D}

To gain more precise information about the role of amplitude $A$ and frequency $\omega$ of the oscillating QDs energy levels  we study here the charge currents averaged over a period $T=2\pi/\omega$ of the driving field. Our main objective is to analyze the spectrum of subgap quasiparticles visible in  nonequilibrium transport properties of the N-DQD-S nanostructure. 
For specific analysis we focus on the Andreev current $\langle j_{N\sigma}(t) \rangle_{t_{0}} 
=\frac{1}{T}\int^{t_{0}+T}_{t_{0}}j_{N\sigma}(t)dt$ induced by the source-drain voltage 
$V_{sd}$ and, in analogy to the preceding section, assuming the periodically driven energy 
levels $\varepsilon_{i\sigma}(t)=A\sin(\omega t)$. From the differential conductance 
$G_{N\sigma}(V_{sd})=\frac{d}{dV_{sd}} \langle j_{N\sigma}(t) \rangle_{t_{0}}$ one can  
infer {\em quasienergies} of the in-gap bound states \cite{Baran-2019}. 

Initially, at $t=0$, the oscillating quantum dot levels $\varepsilon_{i\sigma}(t)$ are imposed simultaneously with the bias voltage $\mu_{N}-\mu_{S}=V_{sd}$, assuming both QDs to be empty. We choose the reference time $t_{0}$ at which the transient effects become negligible. This choice can be quite arbitrary, because safely after forming our N-DQD-S heterostructure 
the time-dependent current oscillates with the same period $T$ as enforced on the energy 
levels (c.f.\ Figs.~\ref{fig: 6}-\ref{fig: 8}). Below we discuss the differential 
conductance $G_{N\sigma}(V_{sd})$ obtained numerically for a few representative sets 
of the model parameters. 
\begin{figure*}[ht]
\centering
\includegraphics[width=1\textwidth]{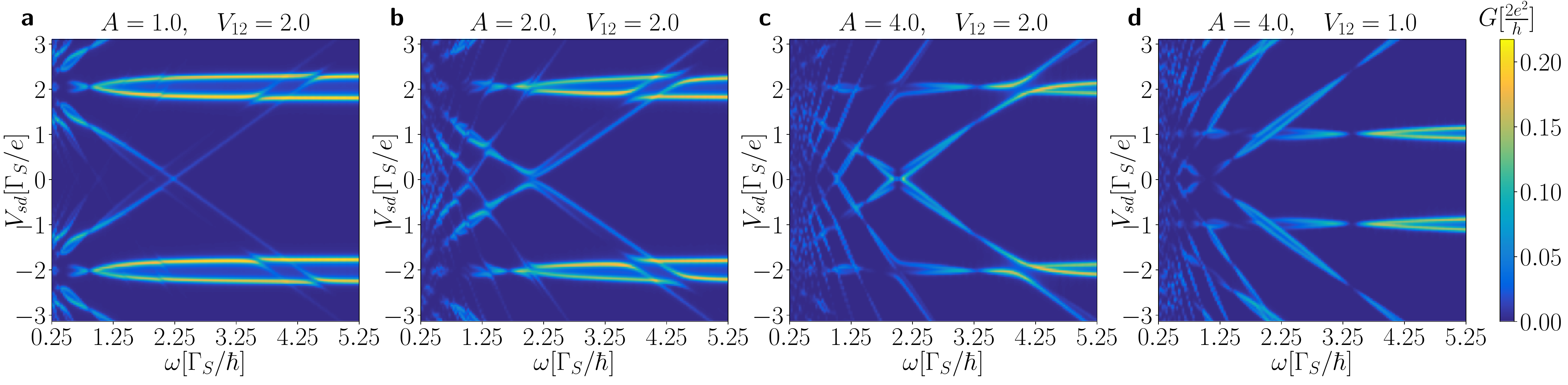}
\caption{{\bf Frequency dependent conductance.} The averaged Andreev conductance 
$G_{N\sigma}(V_{sd})$ in units of  $2e^{2}/h$  as a function 
of the frequency $\omega$ and source-drain voltage $V_{sd}$ obtained for several 
amplitudes $A$ and interdot couplings $V_{12}$ (in units of $\Gamma_{S}$), as indicated. 
We used the model parameters $U=0$, $\Gamma_{N}/\Gamma_{S}=0.1$.}
\label{fig: 9_map}
\end{figure*}

Figure \ref{fig: 9_map} presents the averaged Andreev conductance obtained for two 
values of the interdot coupling $V_{12}$  and several amplitudes $A$, as indicated.
Panels (a-d) display the characteristic features originating from the photon-assisted 
tunneling. We notice that besides the main quasiparticle peaks (for $\Gamma_{N}\ll \Gamma_{S}$) appearing at $\pm\frac{1}{2}\left(\sqrt{4V_{12}^{2}+\Gamma_{S}^{2}/4}
\pm\Gamma_{S}/2\right)$ there emerge  additional side-peaks originating from the stimulated 
emission/absorption of the photon quanta. Their intensity (spectral weight) and 
avoided-crossing behavior are sensitive to the frequency and amplitude of a microwave 
field. The main quasiparticle peaks are replicated at multiples of $\omega$
and they can be interpreted as higher order harmonics of the initial bound states. 

\begin{figure*}[hb]
\centering
\includegraphics[width=1\textwidth]{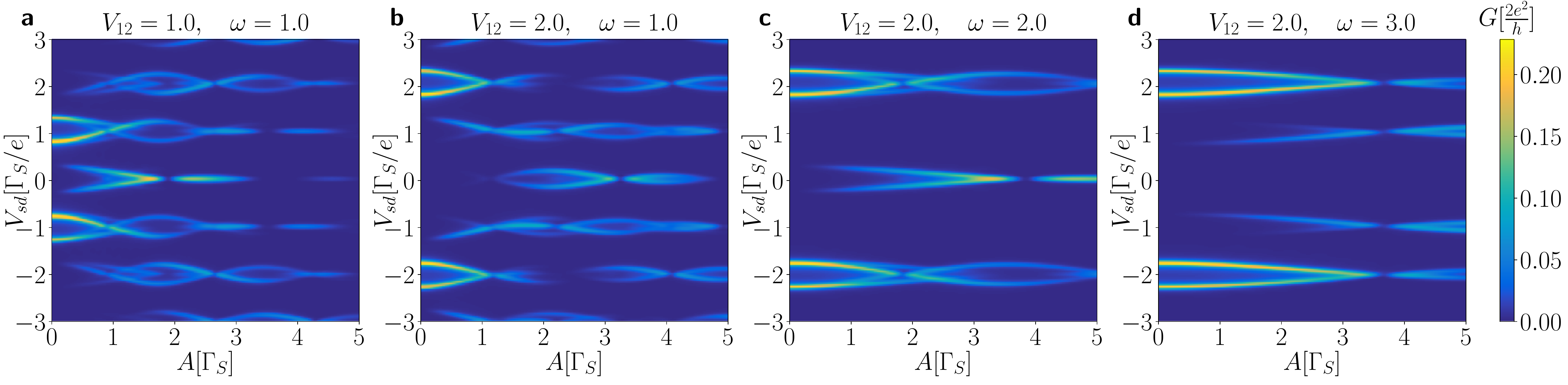}
\caption{{\bf Amplitude dependent conductance.} Variation of the averaged conductance $G_{N\sigma}(V_{sd})$ in units of $2e^{2}/h$ against the amplitude $A$ of the oscillating levels and source-drain voltage $V_{sd}$  obtained for several interdot couplings $V_{12}$ and frequencies $\omega$ (in units of $\Gamma_{S}$), as indicated. Calculations are done for $U=0$ and $\Gamma_{N}/\Gamma_{S}=0.1$.}
\label{fig: 10_map}
\end{figure*}

Basic aspects of the photon-assisted tunneling through the quantum dots sandwiched between the normal electrodes have been extensively studied in literature \cite{Jauho1994,Taranko-2005,Platero-2004}, predicting the main resonance peaks and their $n$-th side-bands modulated by the squared Bessel functions of the first kind $J^{2}_{n}(A/\omega)$. 
As regards the specific photon-assisted tunneling in the superconducting junctions, it has been observed that the differential conductance $G(V_{sd})$ in situations with the single quantum dots can be expressed by $G(V_{sd})=\sum_{n} J^{2}_{n}(kA/\omega) G^{(0)}(V_{sd}+\frac{n\omega}{k})$, where $G^{(0)}(V_{sd})$ corresponds to the conductivity without microwave radiation and $k$ denotes the number of electrons transferred in an elementary tunneling process  \cite{Kot-2020,Franke-2020}. 
For our N-DQD-S nanostrocture we notice that the main resonant peaks and their side-bands are weighted by the squared Bessel function $J_{0}^{2}\left( \frac{2A}{\omega}\right)$. The main resonance peaks and side-bands disappear at such frequencies $\omega$ for which the Bessel function vanishes. Fig.~\ref{fig: 9_map}d shows such points for  $\omega/\Gamma_{S} \sim 3.3$, $1.45$, $0.92$, corresponding to the first, second and third zeros of $J_{0}(2A/\omega)$. For some given amplitude $A$ the frequency $\omega$ at which the main quasiparticle peaks and their higher harmonics disappear is independent of the interdot coupling $V_{12}$ (Figs.~\ref{fig: 9_map}c and \ref{fig: 9_map}d).

Let us now consider variation of the averaged Andreev conductance $G_{N\sigma}$ with respect to ($V_{sd},A$) 
for a few values of the interdot coupling $V_{12}$ (Fig.~\ref{fig: 10_map}). 
In absence of the microwave field, $A=0$, there exist four peaks in the differential 
conductance corresponding to two pairs of in-gap bound states. Upon increasing 
a power of the microwave field (for larger amplitude $A$) 
the main quasiparticle peaks loose some part their intensities (spectral weights) at
expense of their new higher-order replicas. By varying the amplitude $A$ such 
replicas appear in the averaged conductance at $\pm\omega$, $\pm 2\omega$, and so on 
around the main peaks. We can also notice that their spectral weight undergoes 
substantial redistribution. In particular, at certain values of the amplitude $A$ 
the spectral weight of individual harmonics vanishes and then reappears. 

\begin{figure}[ht]
\centering
\includegraphics[width=0.5\columnwidth]{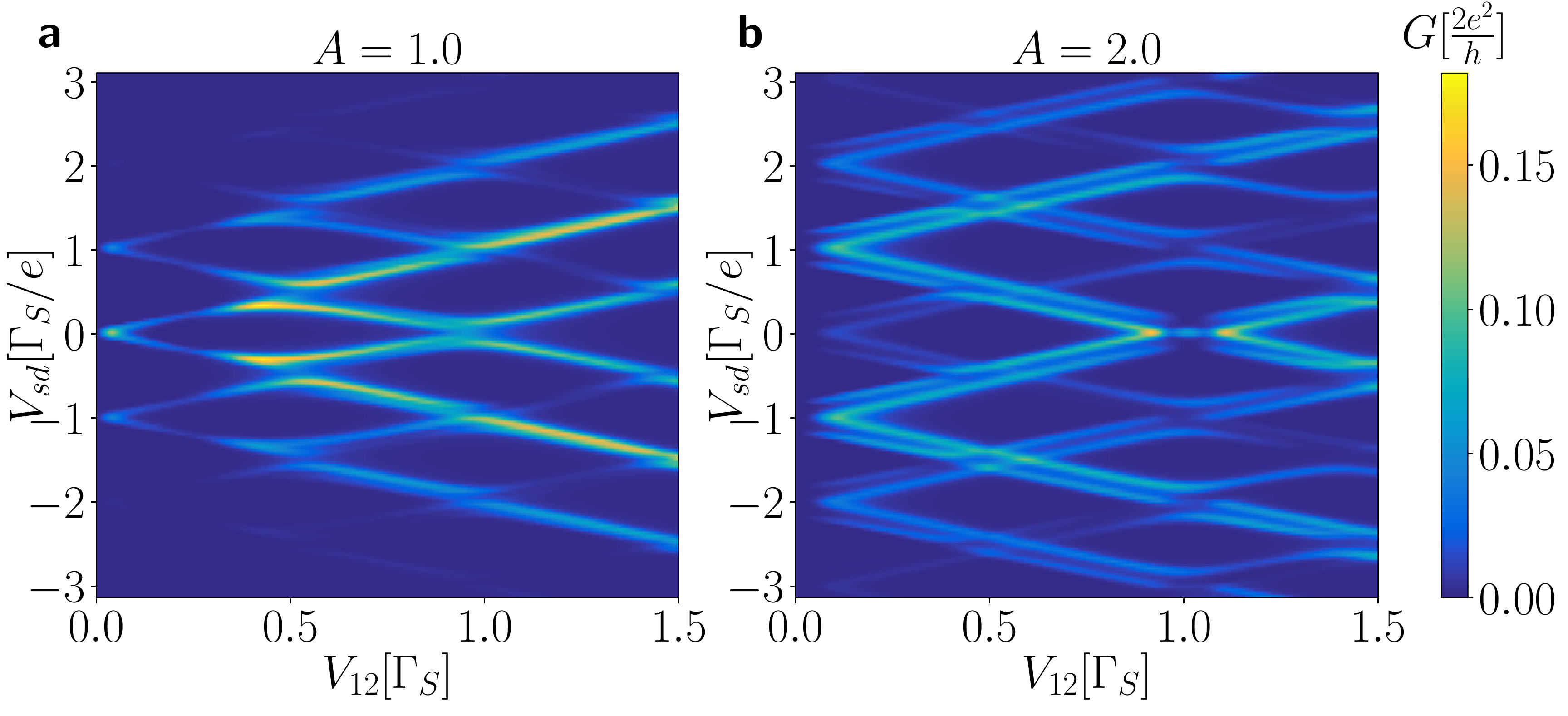}
\caption{{\bf Dependence on interdot coupling.} The averaged Andreev conductance 
$G_{N\sigma}(V_{sd})$ in units of  $2e^{2}/h$  as a function of the interdot coupling $V_{12}$ 
and source-drain voltage $V_{sd}$ (in units of $\Gamma_{S}$) obtained for $U=0$,  $\omega/\Gamma_{S}=1$, 
$\Gamma_{N}/\Gamma_{S}=0.1$,  assuming $A/\Gamma_{S}=1$ (left panel) and $A/\Gamma_{S}=2$ (right panel).}
\label{fig: 11_map}
\end{figure}
To check influence of the inter-dot coupling $V_{12}$ on the averaged Andreev conductance  
we present in Fig.~\ref{fig: 11_map} the results obtained for $\omega/\Gamma_{S}=1$ and two 
amplitudes $A/\Gamma_{S}=1$ and $2$. In the first case the peaks, appearing around $\pm n \omega$, 
gradually split into the lower and upper branches with the increasing coupling $V_{12}$. Yet, 
they never cross each other because of the quantum mechanical interference \cite{Melin-2020}. 
For the larger amplitude, $A/\Gamma_{S}=2$, we clearly notice such avoided-crossing tendency, 
where each harmonic consists of two nearby located peaks. This is an example of the $n$-fold 
fine structure driven in the harmonics, whenever the specific constraint $A/\omega=n$ is encountered.

\begin{figure*}[h]
\centering
\includegraphics[width=0.5\textwidth]{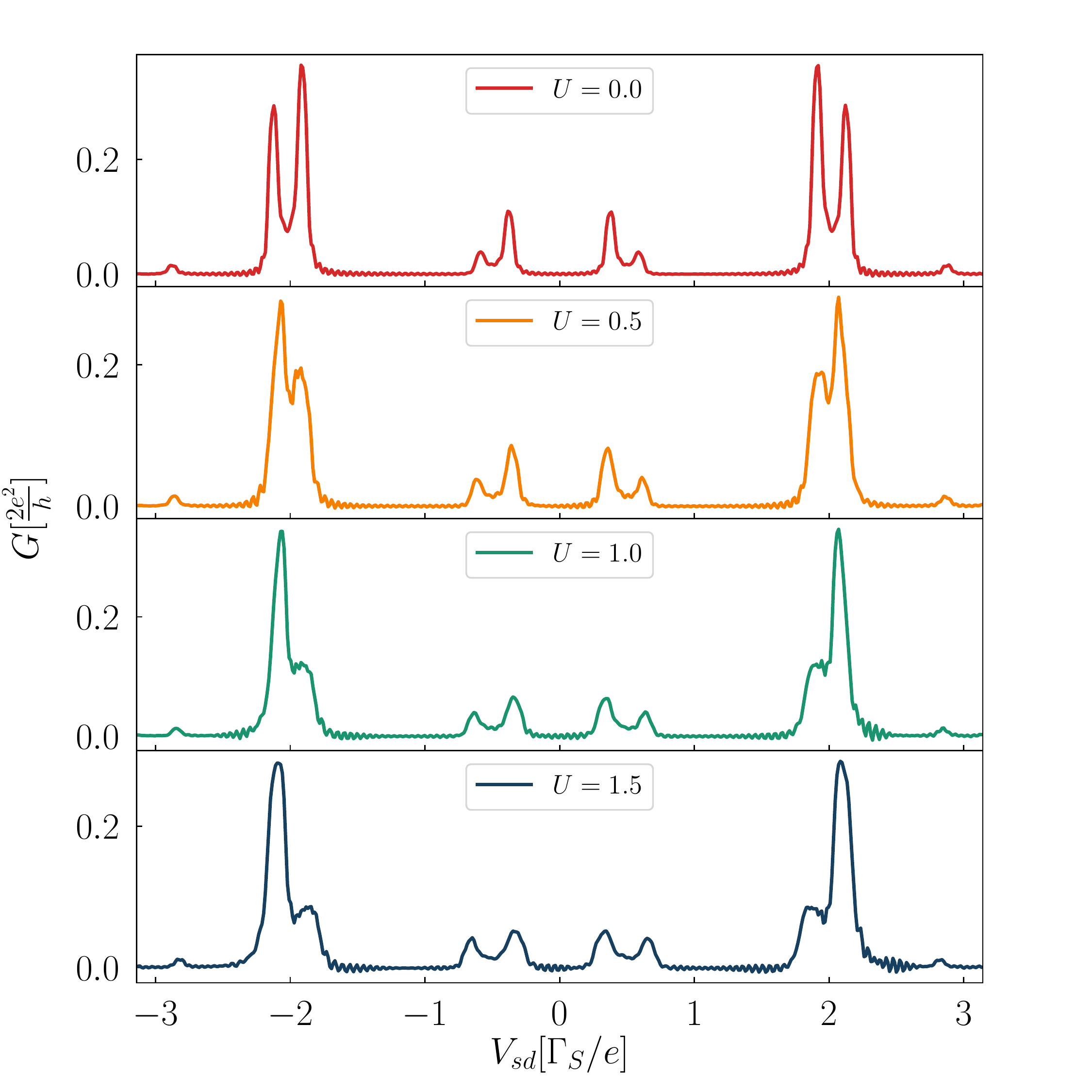}
\caption{{\bf Correlation effects.} The averaged conductance $G_{N\sigma}(V_{sd})$ in units of  $2e^{2}/h$ versus the source-drain voltage $V_{sd}$ obtained within mean-field approximation for several values of $U$  (as indicated), assuming   $A=2$, $V_{12}=2$, $\omega=2.5$, $\Gamma_{N}=0.1$ and  $\Gamma_{S}=1$.}
\label{fig: 12}
\end{figure*}

Finally, in Fig.~\ref{fig: 12} we present the averaged conductance $G_{N\sigma}(V_{sd})$ of the interacting system obtained for $V_{12}/\Gamma_{S}=2$, $A/\Gamma_{S}=2$, $\omega/\Gamma_{S}=2.5$, where panels form top to bottom refer to $U/\Gamma_{S}=0$, $0.5$, $1$,  and $1.5$, respectively. The particle-to-hole scattering mechanism (contributing to the subgap Andreev current) implies the fully symmetric conductance $G_{N\sigma}(-V_{sd})=G_{N\sigma}(V_{sd})$. In the uncorelated system (top panel) the main quasiparticle peaks appear at $\pm\frac{1}{2}\left(\sqrt{4V^2_{12}+\Gamma^{2}_{S}/4}\pm\frac{\Gamma_{S}}{2}\right)$ and their higher order replicas are spaced by $\pm n\omega$. For the presently chosen parameters the second- and higher-order  harmonics become hardly visible because of their very small spectral weights (see Fig.~\ref{fig: 12} and Fig.~\ref{fig: 9_map}b).
Upon increasing the Coulomb potential $U$ the main quasiparticle peaks only slightly change their positions. Major influence of the correlation effects is manifested through noticeable redistribution of the spectral weights, both between the harmonics and between their fine sub-structure. More detailed analysis of the photon-stimulated Andreev transport of the strongly correlated N-DQD-S system would require some sophisticated (nonperturbative) techniques, and such study is beyond the scope of the present work.

In addition to the numerical computations of the averaged current directly from the equations of motion, we have also developed the auxiliary procedure based on machine learning algorithm which reliably yields the Andreev conductance 
for an arbitrary set of the model parameters (see the last subsection of {\em Methods}).

\section*{Discussion}
\label{sec:level5}

We have studied the double quantum dot coupled between the superconducting and normal
leads, addressing its dynamical response to (i) abrupt application 
of the bias voltage, (ii) sudden change of the energy levels, and (iii) their periodic 
driving. These effects can be routinely triggered either by {\em dc} or {\em ac} external 
potentials. We have analyzed the time-dependent charge flow between the external reservoirs 
and the quantum dots, revealing an oscillatory behavior (analogous to the Rabi-type mechanism 
involving pairs of the in-gap quasiparticle states induced by the superconducting proximity 
effect) with a damping caused by the relaxation processes on a continuum spectrum of the normal lead.

Inspecting the time-dependent profiles of various physical observables we have found 
the signatures of such frequency components which coincide with the subgap quasiparticle 
energies. For the quantum quench imposed by the source-drain voltage and by the gate 
potential the dynamics of proximitized double quantum dot reveals superposition of 
the fast and slow oscillatory modes, giving rise to the beating patters. These features 
are well observable over quite long time interval, $\Delta t \sim 10 \hbar/\Gamma_{N}$, 
in contrast to much faster transient phenomena realized in the single quantum dot (N-QD-S) 
heterostructures \cite{LevyYeyati-2017,Taranko-2018}.

In the case of periodically driven energy levels we have found more complex 
time-dependent behavior. Response of the N-DQD-S heterostructure depends both on 
the frequency $\omega$ and amplitude $A$ of the periodically varying levels. We have 
illustrated these phenomena in absence (Figs.\ \ref{fig: 6}-\ref{fig: 8}) and in 
presence of the bias voltage  (Figs.\ \ref{fig: 9_map}-\ref{fig: 11_map}). We 
have predicted that amplitude (related to the power of driving force) has crucial 
effect on activating the higher-order harmonics of in-gap quasiparticle sates, as
evidenced for the unbiased (Fig.\ \ref{fig: 6}) and biased (Fig.\ \ref{fig: 10_map})
heterostructures. The frequency, on the other hand, is manifested by replicas 
of the main quasiparticle peaks. Similar effects have been recently observed 
experimentally in the Josephson-type junctions, comprising the single quantum dot
\cite{Kot-2020,Franke-2020}. In our case the proximitized double quantum dot 
is characterized by a sequence of the photon-assisted enhancements in the differential 
conductance with an additional fine-structure appearing in the harmonics
due to interference effects. Upon varying the frequency (Fig.\ \ref{fig: 9_map})
or the interdot coupling (Fig.\ \ref{fig: 11_map}) the neighboring harmonics
never cross each other because of their quantum mechanical interference, which 
is feasible also in multi-terminal superconducting junctions \cite{Melin-2020}.    

Our considerations could be verified experimentally by means of the subgap 
tunneling spectroscopy using the carbon nanotubes, semiconducting nanowires or other 
lithographically constructed quantum dots embedded  between the superconducting 
and metallic electrodes. Another realization would be possible using the scanning microscope
technique, where the conducting tip can probe the dimerized molecules deposited on
superconducting substrates. The characteristic time-scales determined in this work might
be important for designing logical operations with use of the superconducting qubits  
\cite{Aguado-2020}. In future studies it would be worthwhile to perform more systematic 
consideration of the correlation effects and address the dynamics of topologically 
nontrivial superconducting nanostructures.

\section*{\label{sec:level6}ACKNOWLEDGEMENTS}
This work was supported by the National Science Centre (NCN, Poland) under grants 
UMO-2017/27/B/ST3/01911 (B.B., R.T.) and UMO-2018/29/B/ST3/00937 (T.D.).

\section*{Methods}

{\small
\subsection*{\label{sec: Eigenfunctions}Eigenvalues and eigenfunctions of the proximitized DQD}

The Hilbert space of the DQD proximitized to superconducting lead is spanned by 16 vectors. In the occupancy representation the matrix Hamiltonian has a block structure, consisting of 6 subspaces \cite{Scherubl_2019}. Two 4-dimensional subspaces contain states with odd number of electrons $|$QD$_{2}$,QD$_{1}\rangle \Rightarrow |0,\uparrow\rangle$, $|\uparrow, 0\rangle$, $|\uparrow\downarrow,\uparrow\rangle$, $|\uparrow,\uparrow\downarrow\rangle$ and $|0,\downarrow\rangle$, $|\downarrow, 0\rangle$, $|\uparrow\downarrow,\downarrow\rangle$, $|\downarrow,\uparrow\downarrow\rangle$, respectively. The next two states $|\uparrow, \uparrow\rangle$, $|\downarrow, \downarrow\rangle$ are decoupled from each other. The remaining 6-dimensional subspace contains the states with even number of electrons,  $|0,0\rangle$, $|0, \uparrow\downarrow\rangle$, $|\uparrow\downarrow,0\rangle$, $|\uparrow\downarrow,\uparrow\downarrow\rangle$, $|\uparrow,\downarrow\rangle$ and $|\downarrow,\uparrow\rangle$, respectively. Diagonalizing the effective matrix Hamiltonian, one obtains for $\varepsilon_{i\sigma}=0$ the following set of eigenvalues $E_{i}$ and eigenfunctions $|\phi_{i}\rangle$ 

\begin{center}
\begin{tabular}{ c|c|c } 
 
 $i$ & $E_{i}$ &  $|\phi_{i}\rangle$ \\ 
 \hline
 1/2 & $\pm\varepsilon$ & $a(|0,\uparrow\rangle \pm |\uparrow\downarrow,\uparrow\rangle)+b(\pm |\uparrow,0\rangle + |\uparrow,\uparrow\downarrow\rangle)$ \\ 
 \hline
 3/4 & $\pm\varepsilon\mp\Gamma_{S}/2$ & $b(|0,\uparrow\rangle \mp |\uparrow\downarrow,\uparrow\rangle) +a(\pm|\uparrow,0\rangle - |\uparrow,\uparrow\downarrow\rangle)$ \\ 
 \hline
 5/6 & $\pm\varepsilon$ & $a(|0,\downarrow\rangle \pm |\uparrow\downarrow,\downarrow\rangle)+b(\pm |\downarrow,0\rangle + |\downarrow,\uparrow\downarrow\rangle)$ \\
 \hline
 7/8 & $\pm\varepsilon\mp\Gamma_{S}/2$ & $b(|0,\downarrow\rangle \mp |\uparrow\downarrow,\downarrow\rangle) +a(\pm|\downarrow,0\rangle - |\downarrow,\uparrow\downarrow\rangle)$ \\
 \hline
 9 & 0 & $|\uparrow,\uparrow\rangle$ \\
 \hline
 10 & 0 & $|\downarrow,\downarrow\rangle$ \\
 \hline
 11 & 0 &  $\frac{\sqrt{2}}{\sqrt{4V^{2}_{12}+\Gamma^{2}_{S}/4}}\left( V_{12}(|0,0\rangle+|\uparrow\downarrow,\uparrow\downarrow\rangle)-\frac{\Gamma_{S}}{4}(|\uparrow,\downarrow\rangle+|\downarrow,\uparrow\rangle) \right)$ \\
 \hline
 12 & 0 & $\frac{1}{\sqrt{2}}(|\uparrow,\downarrow\rangle -|\downarrow,\uparrow\rangle)$\\
 \hline
 13/14 & $\pm\Gamma_{S}/2$ & $\frac{1}{2}\left( |0,0\rangle-|\uparrow\downarrow,\uparrow\downarrow\rangle \pm |0,\uparrow\downarrow\rangle \mp |\uparrow\downarrow,0\rangle \right)$ \\
 \hline
 15/16 & $\pm\sqrt{4V^{2}_{12}+\Gamma^{2}_{S}/4}$ & $\frac{\Gamma_{S}}{4\sqrt{4V^{2}_{12}+\Gamma^{2}_{S}/4}}(|0,0\rangle+|\uparrow\downarrow,\uparrow\downarrow\rangle) \pm\frac{1}{2}(|0,\uparrow\downarrow\rangle+|\uparrow\downarrow,0\rangle)  +\frac{V_{12}}{\sqrt{4V^{2}_{12}+\Gamma^{2}_{S}/4}}(|\uparrow,\downarrow\rangle+|\downarrow,\uparrow\rangle)$ \\

\end{tabular}
\end{center}
where $\varepsilon=\frac{1}{2}\left(\sqrt{4V^{2}_{12}+\Gamma^{2}_{S}/4}+\Gamma_{S}/2\right)$, $a=\frac{1}{\sqrt{2}}\frac{V_{12}}{\sqrt{V^{2}_{12}+\varepsilon^{2}}}$ and $b=\frac{1}{\sqrt{2}}\frac{\varepsilon}{\sqrt{V^{2}_{12}+\varepsilon^{2}}}$.

{\small
\subsection*{\label{sec: EOM}Equations of motion}

Here, we explicitly present the set of differential equations needed for determination of
the time-dependent occupancy $n_{i\sigma}(t)=\langle \hat{c}^{\dagger}_{i\sigma}(t)\hat{c}_{i\sigma}(t)\rangle$ 
and other functions coupled to it. Using the exact formula 
\begin{equation}
\hat{c}_{N\textbf{k}\sigma}(t) = \hat{c}_{N\textbf{k}\sigma}(0)\exp{\left( -i\int^{t}_{0}dt'\varepsilon_{N\textbf{k}\sigma}(t')\right) } -i\int^{t}_{0}dt'\hat{c}_{2\sigma}(t') V_{N\textbf{k}}\exp{\left( -i\int^{t}_{t'}d\tau\varepsilon_{N\textbf{k}\sigma}(\tau)\right) }
\label{eq: A0}
\end{equation}
and applying the wide band limit approximation we derive the following set of equations 
\begin{eqnarray}
\frac{dn_{1\sigma}(t)}{dt} &=&2 \textrm{Im}(V_{12}\langle \hat{c}^{\dagger}_{1\sigma}(t)\hat{c}_{2\sigma}(t)\rangle -\frac{\Gamma_{S}}{2}\langle \hat{c}_{1-\sigma}(t)\hat{c}_{1\sigma}(t)\rangle),
\label{eq: A1}
\\
\frac{dn_{2\sigma}(t)}{dt}&=&2\textrm{Im}[-V_{12}\langle \hat{c}^{\dagger}_{1\sigma}(t)\hat{c}_{2\sigma}(t)\rangle-\frac{i\Gamma_{N}}{2}n_{2\sigma}(t)+\sum_{\textbf{k}}V_{N\textbf{k}}\exp(-i\varepsilon_{N\textbf{k}}t)\langle \hat{c}^{\dagger}_{2\sigma}(t)\hat{c}_{N\textbf{k}\sigma}(0)\rangle\beta],
\label{eq: A2}
\\
\frac{d\langle \hat{c}_{1\sigma}(t)\hat{c}_{2-\sigma}(t)\rangle}{dt}&=&\left[-i\left(\varepsilon_{1\sigma}+\varepsilon_{2-\sigma}\right)-\frac{\Gamma_{N}}{2} \right]\langle \hat{c}_{1\sigma}(t)\hat{c}_{2-\sigma}(t)\rangle -iV_{12}\left( \langle \hat{c}_{1\sigma}(t)\hat{c}_{1-\sigma}(t)\rangle+ \langle \hat{c}_{2\sigma}(t)\hat{c}_{2-\sigma}(t)\rangle\right)\nonumber\\
&&+ \alpha i \frac{\Gamma_{S}}{2}\langle \hat{c}^{\dagger}_{1-\sigma}(t)\hat{c}_{2-\sigma}(t) \rangle-i\sum_{\textbf{k}}V_{N\textbf{k}}\exp(-i\varepsilon_{N\textbf{k}}t)\langle \hat{c}_{1\sigma}(t)\hat{c}_{N\textbf{k}-\sigma}(0)\rangle\beta ,
\label{eq: A3}
\\
\frac{d\langle \hat{c}_{1\downarrow}(t)\hat{c}_{1\uparrow}(t)\rangle}{dt}&=&-i\left(\varepsilon_{1\uparrow}+\varepsilon_{1\downarrow}\right)\langle \hat{c}_{1\downarrow}(t)\hat{c}_{1\uparrow}(t)\rangle -iV_{12}\left( \langle \hat{c}_{1\downarrow}(t)\hat{c}_{2\uparrow}(t)\rangle- \langle \hat{c}_{1\uparrow}(t)\hat{c}_{2\downarrow}(t)\rangle\right)- i \frac{\Gamma_{S}}{2}\left(1-\sum_{\sigma}n_{1\sigma}(t) \right),
\label{eq: A4}
\\
\frac{d\langle \hat{c}_{2\downarrow}(t)\hat{c}_{2\uparrow}(t)\rangle}{dt}&=&\left[-i\left(\varepsilon_{2\uparrow}+\varepsilon_{2\downarrow}\right)-\Gamma_{N}\right]\langle \hat{c}_{2\downarrow}(t)\hat{c}_{2\uparrow}(t)\rangle +iV_{12}\left( \langle \hat{c}_{1\uparrow}(t)\hat{c}_{2\downarrow}(t)\rangle- \langle \hat{c}_{1\downarrow}(t)\hat{c}_{2\uparrow}(t)\rangle\right)\nonumber\\
&&+ i\sum_{\textbf{k}}V_{N\textbf{k}}\exp(-i\varepsilon_{N\textbf{k}}t)\left( \langle \hat{c}_{2\uparrow}(t)\hat{c}_{N\textbf{k}\downarrow}(0) \rangle - \langle \hat{c}_{2\downarrow}(t)\hat{c}_{N\textbf{k}\uparrow}(0) \rangle\right)\beta,
\label{eq: A5}
\\
\frac{d\langle \hat{c}^{\dagger}_{1\sigma}(t)\hat{c}_{2\sigma}(t)\rangle}{dt}&=&\left[-i\left(\varepsilon_{2\sigma}-\varepsilon_{1\sigma}\right)-\frac{\Gamma_{N}}{2} \right]\langle \hat{c}^{\dagger}_{1\sigma}(t)\hat{c}_{2\sigma}(t)\rangle -iV_{12}\left(n_{1\sigma}(t)-n_{2\sigma}(t)\right)+\alpha i \frac{\Gamma_{S}}{2}\langle \hat{c}_{1-\sigma}(t)\hat{c}_{2\sigma}(t) \rangle\nonumber\\
&&- i\sum_{\textbf{k}}V_{N\textbf{k}}\exp(-i\varepsilon_{N\textbf{k}}t)\langle \hat{c}^{\dagger}_{1\sigma}(t)\hat{c}_{N\textbf{k}\sigma}(0)\rangle\beta,
\label{eq: A6}
\end{eqnarray}
where $\alpha=+(-)$, $\beta = \exp(-i(t-t_{1})V_{sd})$, $t_{1}$ denotes the time at which the bias voltage $V_{sd}$ is applied and $\langle \dots\rangle$ stands for the quantum statistical averaging. At this level there appear the new correlation functions $\langle \hat{A}_{i\sigma}(t)\hat{B}_{\textbf{k}\sigma}(0)\rangle$, where $\hat{A}$ ($\hat{B}$) corresponds to the creation or annihilation operator of electron in the quantum dots (the normal lead). These functions can be determined from the the following equations of motion
\begin{eqnarray}
\frac{d\langle \hat{c}^{\dagger}_{1\sigma}(t)\hat{c}_{N\textbf{k}\sigma}(0)\rangle}{dt} &=& i \varepsilon_{1\sigma}\langle \hat{c}^{\dagger}_{1\sigma}(t)\hat{c}_{N\textbf{k}\sigma}(0) \rangle +iV_{12}\langle \hat{c}^{\dagger}_{2\sigma}(t)\hat{c}_{N\textbf{k}\sigma}(0)\rangle +\alpha i \frac{\Gamma_{S}}{2}\langle \hat{c}_{1-\sigma}(t)\hat{c}_{N\textbf{k}\sigma}(0)\rangle,
\label{eq: A7}
\\
\frac{d\langle \hat{c}_{1\sigma}(t)\hat{c}_{N\textbf{k}-\sigma}(0)\rangle}{dt} &=& -i \varepsilon_{1\sigma}\langle \hat{c}_{1\sigma}(t)\hat{c}_{N\textbf{k}-\sigma}(0) \rangle -iV_{12}\langle \hat{c}_{2\sigma}(t)\hat{c}_{N\textbf{k}-\sigma}(0)\rangle-\alpha i \frac{\Gamma_{S}}{2}\langle \hat{c}^{\dagger}_{1-\sigma}(t)\hat{c}_{N\textbf{k}-\sigma}(0)\rangle,
\label{eq: A8}
\\
\frac{d\langle \hat{c}^{\dagger}_{2\sigma}(t)\hat{c}_{N\textbf{k}\sigma}(0)\rangle}{dt} &=& \left(i \varepsilon_{2\sigma}-\frac{\Gamma_{N}}{2}\right)\langle \hat{c}^{\dagger}_{2\sigma}(t)\hat{c}_{N\textbf{k}\sigma}(0) \rangle  +iV_{12}\langle \hat{c}^{\dagger}_{1\sigma}(t)\hat{c}_{N\textbf{k}\sigma}(0)\rangle+iV_{N\textbf{k}}e^{i\varepsilon_{N\textbf{k}}t}\langle \hat{n}_{\textbf{k}\sigma}(0)\rangle\beta^{-1},
\label{eq: A9}
\\
\frac{d\langle \hat{c}_{2\sigma}(t)\hat{c}_{N\textbf{k}-\sigma}(0)\rangle}{dt} &=& \left(-i \varepsilon_{2\sigma}-\frac{\Gamma_{N}}{2}\right)\langle \hat{c}_{2\sigma}(t)\hat{c}_{N\textbf{k}-\sigma}(0) \rangle-iV_{12}\langle \hat{c}_{1\sigma}(t)\hat{c}_{N\textbf{k}-\sigma}(0)\rangle,
\label{eq: A10}
\end{eqnarray}
where $\langle \hat{n}_{\textbf{k}\sigma}(0)\rangle = \left[1 +\mbox{\rm exp}\left( (\varepsilon_{N{\bf k}\sigma}-\mu_{N})/k_{B}T\right) \right]^{-1}$ is the Fermi distribution function for the normal lead electrons.

We have solved numerically these coupled differential equations (\ref{eq: A1}-\ref{eq: A10}) subject to the specific initial 
conditions. For convenience, we have assumed that at $t = 0$ both external reservoirs were isolated from the quantum dots. 
In next steps, we have calculated iteratively the time-dependent observables using the Runge Kutta algorithm with sufficiently
dense equidistant temporal points $t \rightarrow t + \delta t \rightarrow  ... \rightarrow t + N \delta t \equiv t_{f}$.
}

\small{
\subsection*{Machine learning approach}
\label{app:B}

Results presented in the main part of this paper have been obtained by solving the differential equations derived for N-DQD-S 
heterostructure. The computational procedure has been rather straightforward (see the preceding section), but required 
quite a lot of time and resources. For instance to produce the conduction maps (Figs.\ \ref{fig: 9_map}-\ref{fig: 11_map}) 
with $150 \times 150$ points resolution it takes approximately one week performing multiprocessing calculations on 
CPU 2x Xeon E5-2660 2.2GHz 16 cores/ 32 threads. This problem motivated us to construct a machine learning model for 
our system. 

To train our neural network we have used the collected set of data of 76 different conductance maps (with different 
resolutions), giving us 971760 conductance data points. Subsequently, we have linearly interpolated every single map 
to doubly increase a number of the data points, finally giving us 3887040 data points. For this purpose we have used 
the open-source software for machine learning - Tensorflow with application programming interface - Keras. 

This neural network has a character of the densely connected type, with 4 input parameters ($V_{12}$,$\omega$,$V_{sd}$,$A$) 
describing noninteracting N-DQD-S setup  and 1 single neuron on the output, specifying the averaged Andreev conductance $G_{N\sigma}$. 
The neural network is composed of 4 hidden layers consisting of 2048, 1024, 512, 256 neurons, respectively. Every hidden 
layer has a dropout of $1\%$ neurons (which helps to avoid over-fitting our model) and, as an activation function, we 
have used sigmoid function. One can notice that this neural network is large, because of non-linearity in the system. 
To train our neural network we have chosen $batch=1024$ and $epoch=600$, giving us the fidelity coefficient 
$R^2=0.987$. Fig.~\ref{fig: ml_test} compares the calculated $G_{N\sigma}$ with respect to the value predicted 
by our neural network. Predictive strength of the machine learning algorithm is illustrated in Fig.~\ref{fig: ml_map}, which shows the conductance maps obtained from the direct calculation (panel a) and by the neural network (panel b) for such model parameters which were not used during the training process.
This neural network model of N-DQD-S heterostructure is available at the following
\href{https://www.dropbox.com/sh/0hzs9im3d3bf0jr/AADRr3kltw2mOdCCh8tedoIWa?dl=0}{www.dropbox.com/sh/0hzs9im3d3bf0jr/AADRr3kltw2mOdCCh8tedoIWa?dl=0} webpage.

\begin{figure}[H]
\centering
\includegraphics[width=0.75\textwidth]{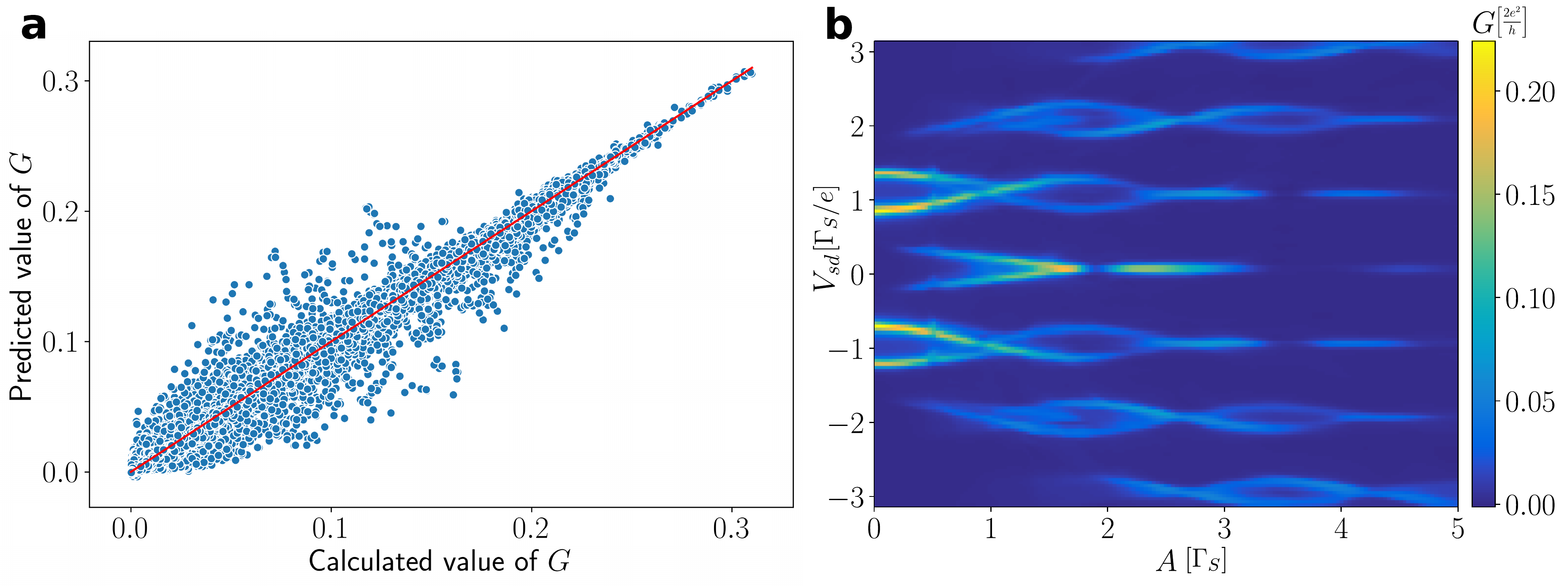}
\caption{{\bf Neural network data.} a) Comparison of the differential conductance predicted by the neural network versus its value determined 
by the microscopic calculations. The red line $y=x$ is a guide to eye. b) The conductance map generated by the neural 
network, reproducing the results presented in  Fig.\ref{fig: 10_map}a.}
\label{fig: ml_test}
\end{figure}
\begin{figure}[H]
\centering
\includegraphics[width=0.75\textwidth]{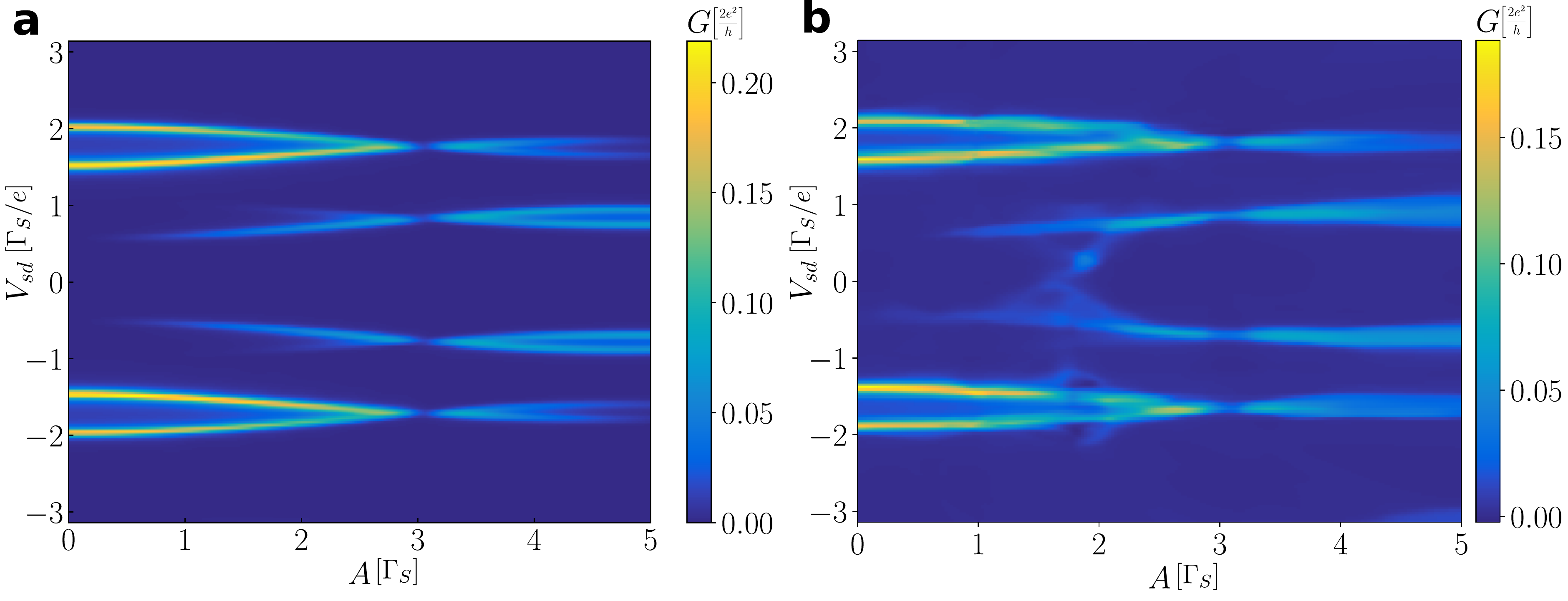}
\caption{{\bf Machine learning results.} The conductance map obtained from the microscopic numerical calculations (panel a) and generated 
by the neural network (panel b) for $V_{12}=1.7$, $\omega=2.5$. The map shown in panel a has not been used 
for learning the neural network.}
\label{fig: ml_map}
\end{figure}
}

\section*{Author contributions statement}
B.B. performed the numerical calculations, R.T. provided the methodological
instruction, and T.D. coordinated this research project. All authors discussed 
the results and prepared the manuscript.

\section*{Additional information}

\textbf{Competing interests} 
The authors declare no competing interests.

\bibliography{biblio.bib}
\end{document}